\documentclass[11pt,dvips]{article}

\usepackage{amsmath,latexsym,amssymb}
\usepackage[dvips]{graphicx}

\newcommand{\f}[2]{\frac{ #1}{ #2}}

\newcommand{\la}{\langle}
\newcommand{\ra}{\rangle}
\newcommand{\lla}{\la\!\la}
\newcommand{\rra}{\ra\!\ra}
\newcommand{\Tr}{{\rm Tr}\,}

\newcommand{\C}{{\cal C}}
\newcommand{\W}{{\cal W}}

\renewcommand{\Re}{{\rm Re\,}}
\renewcommand{\Im}{{\rm Im\,}}

\newcommand{\nsection}[1]{\section{#1}
\setcounter{equation}{0}}

\begin{document}

{\thispagestyle{empty}
\vfil
\noindent \hspace{1cm} \hfill Revised version 

\noindent \hspace{1cm} \hfill March 2014 

\begin{center}
  \vspace*{1.0cm}
  {\large\bf HADRONIC TOTAL CROSS SECTIONS AT HIGH ENERGY AND THE QCD
    SPECTRUM}
  \\
  \vspace*{1.0cm}
  {\large Matteo Giordano\footnote{E--mail: giordano@atomki.mta.hu} and
    Enrico Meggiolaro\footnote{E--mail: enrico.meggiolaro@df.unipi.it}}\\
  \vspace*{0.5cm}{\normalsize
    {$^{1}$ Institute for Nuclear Research of the Hungarian Academy of
      Sciences (ATOMKI),\\
      Bem t\'er 18/c, H--4026 Debrecen, Hungary}}\\
  \vspace*{0.5cm}{\normalsize
    {$^{2}$ Dipartimento di Fisica, Universit\`a di Pisa,
      and INFN, Sezione di Pisa,\\
      Largo Pontecorvo 3, I--56127 Pisa, Italy}}\\
  \vspace*{2cm}{\large \bf Abstract}
\end{center}

\noindent
We show how to obtain the leading energy dependence of hadronic total
cross sections, in the framework of the nonperturbative approach to
{\it soft} high-energy scattering based on Wilson-loop correlation
functions, if certain nontrivial analyticity assumptions are
satisfied. The total cross sections turn out to be of ``Froissart''
type, $\sigma_{\rm tot}^{(hh)}(s) \mathop\sim B\log^2 s$ for ${s \to
  \infty}$. We also discuss under which conditions the coefficient $B$
is universal, i.e., independent of the hadrons involved in the
scattering process. In the most natural scenarios for universality,
$B$ can be related to the stable spectrum of QCD, and is predicted to
be $B_{\rm th}\simeq 0.22~{\rm mb}$, in fair agreement with
experimental results. If we consider, instead, the stable spectrum of
the {\it quenched} (i.e., pure-gauge) theory, we obtain a quite larger
value $B^{(Q)}_{\rm th} \ge 0.42~{\rm mb}$, suggesting (quite
surprisingly) large {\it unquenching} effects due to the {\it sea}
quarks.
\\
\addtocounter{footnote}{-2}
\vfil}
\newpage

\nsection{Introduction}
\label{sec:0}

The recent measurements of proton-proton total cross sections at the
LHC by the TOTEM collaboration~\cite{TOTEM1,TOTEM2,TOTEM3,TOTEM4}, at total center-of-mass
energy $\sqrt{s}=7~{\rm TeV}$ and $\sqrt{s}=8~{\rm TeV}$,  have helped
in reviving the study of the high-energy behaviour of hadronic total
cross sections~\cite{KFK,Dremin,Dremin2,GLM}. The main theoretical problem in this
context is providing a convincing explanation of the rise with energy
of total cross sections observed in experiments, and a definite
prediction of its functional form, in the framework of QCD. Despite
many years of efforts, a satisfactory solution to this problem is
still lacking.

Experimental data for total cross sections are well described by a
``Froissart-like'' behaviour $\sigma_{\rm tot}^{(hh)}(s)\mathop\sim
B\log^2 s$ for ${s \to \infty}$, with a {\it universal} (i.e., {\it
  not} depending on the particular hadrons involved) coefficient $B
\simeq 0.25$ -- $0.3$ mb
\cite{Blogs1,Blogs2,Blogs3,Blogs4,Blogs5,Blogs6,Blogs7,Blogs8,pdg}. The
attribute 
``Froissart-like'' is a reference to the functional form appearing in 
the well-known Froissart-\L ukaszuk-Martin (FLM)
bound~\cite{FLM1,FLM2,FLM3}, which states that for $s \to \infty$,
$\sigma^{(hh)}_{\rm tot}(s) \le {\frac{\pi}{m_\pi^2}}\log^2
\left({\frac{s}{s_0}} \right)$, where $m_\pi$ is the pion mass and
$s_0$ is an unspecified squared mass scale.\footnote{Notice that the
  {\it experimental} value of $B$ is much smaller than the coefficient
  ${\frac{\pi}{m_\pi^2}}$ (about $0.5 \%$) appearing in the FLM
  bound. See Refs.~\cite{Martin,WMRS,MR} for recent work to improve
  the bound, and also Ref.~\cite{GdR}.} 
Theoretical supports to this functional form and to the universality
of the coefficient $B$ were found in the model of the iteration of
{\it soft-Pomeron} exchanges by eikonal
unitarisation~\cite{soft-pomeron1,soft-pomeron2} (recently revisited
in the context of holographic QCD~\cite{BKYZ}), and also using
arguments based on the so-called {\it Color Glass Condensate} of
QCD~\cite{CGC1,CGC2}, or simply modifying the original {\it Heisenberg's
  model}~\cite{Heisenberg} in connection with the presence of {\it
  glueballs}~\cite{DGN}. These arguments however do not provide a full
derivation of the ``Froissart-like'' total cross sections from the
first principles of QCD. We mention at this point that the $\log^2 s$
behaviour of total cross sections has been recently questioned in
Refs.~\cite{FMS1,FMS2,FMS3}, and the validity itself of the FLM bound has also
been put under scrutiny~\cite{Azimov} (see however also Ref.~\cite{MR}
for a comment).  

Explaining the behaviour of hadronic total cross sections
is part of the more general problem of hadronic {\it soft} high-energy
scattering, i.e., high-energy elastic scattering of hadrons at low
transferred momentum. {\it Soft} high-energy processes are
characterised by two different energy scales, provided by the total
center-of-mass energy squared $s$, which is large, and the transferred
momentum squared $t$, which is fixed and smaller than the typical
(squared) energy scale of strong interactions ($|t| \lesssim 1~ {\rm
  GeV}^2 \ll s$). As a consequence, the study of these processes
cannot fully rely on perturbation theory. A nonperturbative approach
to this problem in the framework of QCD has been proposed in
Ref.~\cite{Nachtmann91}, and has been further developed in a number of 
papers (see, e.g., Ref.~\cite{pomeron-book} for a review and a complete
list of references): using a functional integral approach, high-energy
hadron-hadron elastic scattering amplitudes are shown to be governed
by the correlation function of certain Wilson loops defined in
Minkowski space~\cite{DFK,Nachtmann97,BN,Dosch,LLCM1}. Moreover, it
has been shown in
Refs.~\cite{analytic1,analytic2,analytic3,Meggiolaro05,GM2009} that
this correlation function can be reconstructed by \emph{analytic 
  continuation} from the correlation function of two Euclidean Wilson
loops. This has allowed the investigation of the correlators using the
nonperturbative methods of Euclidean Field Theory, both through
approximate analytical calculations in the \emph{Stochastic Vacuum
  Model} (SVM)~\cite{LLCM2}, in the \emph{Instanton Liquid Model} 
(ILM)~\cite{ILM,GM2010}, and using the AdS/CFT
\emph{correspondence}~\cite{JP,JP2,Janik,GP2010}, and through
numerical Monte Carlo simulations in \emph{Lattice Gauge Theory}
(LGT)~\cite{GM2008,GM2010} (see also Refs.~\cite{GM2011a,GM2011b} for a short 
review).  

As discussed in Refs.~\cite{GM2008,GM2010}, the comparison of the
analytic nonperturbative calculations in QCD-related
models~\cite{LLCM2,ILM,GM2010} (as well as that of the perturbative
calculations~\cite{BB,Meggiolaro05,LLCM2}) to the numerical data from
LGT is not satisfactory. As the numerical results obtained on the
lattice can be considered ``exact'' (within the errors) predictions of
QCD, this casts doubts on the viability of the above-mentioned models,
which moreover do not lead to rising total cross sections. 

Recently, a new analysis of the numerical results has been proposed in
Ref.~\cite{GMM}. The main purpose of that paper was to provide a
parameterisation of the lattice data that, after analytic continuation
to Minkowski space, results into a physically acceptable scattering
amplitude satisfying the unitarity constraint, and that furthermore
leads to a rising behaviour of total cross sections at high energy
(beside, of course, fitting well the data). In particular, we were
able to identify and qualitatively justify a class of simple
parameterisations that lead to universal ``Froissart-like''
behaviour. Moreover, the value of $B$ resulting from our fits was
of the same order of magnitude of the experimental value, within the
large errors, and notwithstanding the use of the quenched
approximation in the numerical simulations. However, although the 
results above look promising, the functional forms used in the
analysis of Ref.~\cite{GMM} are not fully justified.   

The purpose of this paper is to gain more insight both on the
functional form of the relevant Wilson-loop correlators, and on the
quantitative identification of its relevant parameters. The basic idea
is to analyse the Euclidean correlators by inserting a complete set of
states between the Wilson loops, and extracting the large
impact-parameter behaviour of the Wilson-loop correlator. Under the 
assumption that the analytic continuation to Minkowski space can be
performed term by term, we are able to identify the terms that
dominate the sum at high energy, and in turn to compute the
high-energy behaviour of total cross sections. Under the
above-mentioned nontrivial analyticity assumption, we provide a
derivation of the ``Froissart-like'' behaviour of hadronic total cross 
sections in the framework of QCD. Furthermore, we discuss how one can
obtain universality of this behaviour, and how the coefficient $B$ of
the $\log^2s$ term is related to the hadronic spectrum. 

The plan of the paper is the following. In Section \ref{sec:1} we give
a brief account of the nonperturbative approach to {\it soft}
high-energy scattering, based on the correlation function of Wilson
loops in the sense of the QCD functional integral. We also discuss the
issue of analytic continuation to Euclidean space. In Section
\ref{sec:outline} we give a general outline of our argument, to
provide a guide for the reader to the more technical discussion of the
following Sections. In Section \ref{sec:2} we relate the
functional-integral language with the operator language, and we
re-express the Wilson-loop correlation function in terms of a sum over
a complete set of states. After performing the analytic continuation
to Minkowski space, we investigate the limits of large energy and
large impact parameter. In Section \ref{sec:3} we use the
corresponding results to investigate the high-energy behaviour of the
hadronic total cross sections and of the elastic scattering
amplitudes. Finally, in Section \ref{sec:c} we draw our
conclusions. Some technical details are discussed in Appendix \ref{app:1}.

\nsection{Meson-meson scattering from dipole-dipole scattering}
\label{sec:1}

In this Section we briefly sketch the nonperturbative approach to {\it
  soft} high-energy scattering (see Ref.~\cite{GM2008} for a more detailed
presentation, and also Ref.~\cite{reggeon} for a recent re-derivation of
the main formula). The elastic scattering amplitude ${\cal M}_{(hh)}$ of 
two hadrons, or more precisely {\it mesons} (taken for simplicity with
the same mass $m$), in the {\it soft} high-energy regime can be
reconstructed from the scattering amplitude ${\cal M}_{(dd)}$ of two
dipoles of fixed transverse sizes $\vec{R}_{1,2\perp}$, and fixed
longitudinal-momentum fractions $f_{1,2}$ of the two quarks in the two
dipoles, after folding with two proper squared hadron wave functions
$|\psi_1|^2$ and $|\psi_2|^2$, describing the two 
interacting hadrons~\cite{DFK,Nachtmann97,BN,Dosch,LLCM1}: 
\begin{equation}
  \label{eq:scatt-hadron}
  \begin{aligned}
{\cal M}_{(hh)}(s,t) =&
\displaystyle\int d^2\vec{R}_{1\perp} \int_0^1 df_1~
|\psi_1(\vec{R}_{1\perp},f_1)|^2
\displaystyle\int d^2\vec{R}_{2\perp} \int_0^1 df_2~
|\psi_2(\vec{R}_{2\perp},f_2)|^2
 \\
&\times {\cal M}_{(dd)}(s,t;\vec{R}_{1\perp},f_1,\vec{R}_{2\perp},f_2)
\equiv \lla {\cal M}_{(dd)}(s,t;\nu_1,\nu_2) \rra ,
  \end{aligned}
\end{equation}
with: $\int d^2\vec{R}_{1\perp} \int_0^1 df_1~
|\psi_1(\vec{R}_{1\perp},f_1)|^2 =
\int d^2\vec{R}_{2\perp} \int_0^1 df_2~
|\psi_2(\vec{R}_{2\perp},f_2)|^2 = 1$, so that $\lla 1\rra =1$. The
notation $\nu_i=(\vec{R}_{i\perp},f_i)$ will be often used for the
sake of brevity. For the treatment of baryons, a similar picture can
be adopted, using a genuine three-body configuration or, alternatively
and even more simply, a quark-diquark configuration: we refer the
interested reader to the above-mentioned original
references~\cite{DFK,Nachtmann97,BN,Dosch,LLCM1} and to
Ref.~\cite{DR}. 

\begin{figure}[t]
  \centering
  \includegraphics[width=0.5\textwidth]{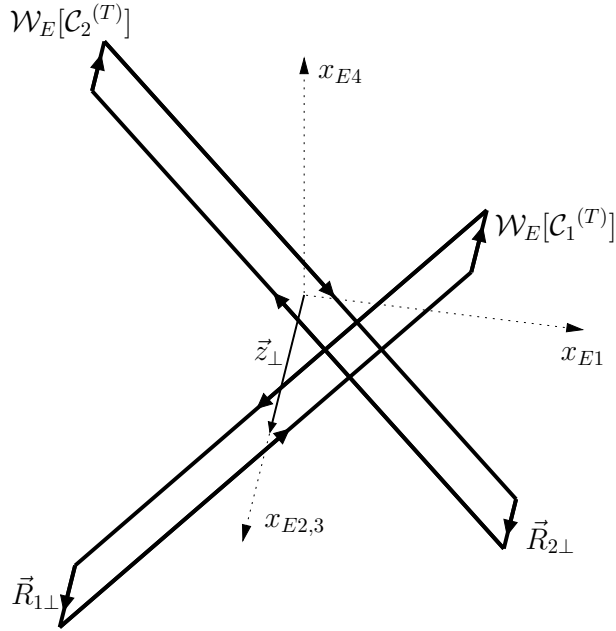}
  \caption{The relevant Wilson loops in Euclidean space.}
  \label{fig:1}
\end{figure}

In turn, the dipole-dipole scattering amplitude is obtained from the
(properly normalised) correlation function of two Wilson loops in the
fundamental representation, defined in Minkowski spacetime, running
along the paths made up of the quark and antiquark classical
straight-line trajectories, and thus forming a hyperbolic angle $\chi
\simeq \log(s/m^2)$ in the longitudinal plane. The paths
are cut at proper times $\pm T$ as an infrared regularisation, and
closed by straight-line ``links'' in the transverse plane, in order to
ensure gauge invariance. Eventually, the limit $T\to\infty$ has to be
taken. It has been shown in
Refs.~\cite{analytic1,analytic2,analytic3,Meggiolaro05,GM2009} 
that the relevant Minkowskian correlation function ${\cal
  G}_M(\chi;T;\vec{z}_\perp;\nu_1,\nu_2)$ ($\vec{z}_\perp$ being the
{\it impact parameter}, i.e., the transverse separation between the
two dipoles) can be reconstructed, by means of {\it analytic
  continuation}, from the Euclidean correlation function of two
Euclidean Wilson loops, 
\begin{equation}
\label{GE}
{\cal G}_E(\theta;T;\vec{z}_\perp;\nu_1,\nu_2) \equiv
\dfrac{\langle {\cal W}_E[{\cal C}^{\,(T)}_1]{\cal W}_E[{\cal C}^{\,(T)}_2] 
\rangle_E}{\langle {\cal W}_E[{\cal C}^{\,(T)}_1]  \rangle_E
\langle {\cal W}_E[{\cal C}^{\,(T)}_2] \rangle_E } - 1\,, 
\end{equation}
where $\langle\ldots\rangle_E$ is the average in the sense of the
Euclidean QCD functional integral. The Euclidean Wilson loop is
defined as follows, 
\begin{equation}
  \label{eq:EWL}
  {\cal W}_E[{\cal C}] \equiv 
{\displaystyle\dfrac{1}{N_c}} \Tr P\!\exp
\left\{ -ig \displaystyle\oint_{{\cal C}}
 {A}_{E\mu}(x_E) dx_{E\mu} \right\}\,,
\end{equation}
where $P$ stands for path-ordering with larger values of the path
parameter appearing {\it on the left}.\footnote{Usually, path-ordering
requires larger values of the path parameter to appear on the right,
while our definition of path-ordering is usually called time-ordering
and is denoted with $T$. The usual convention has been followed in our
previous papers. However, here we will also use the time-ordered
product of operators, for which we have preferred to reserve the
symbol $T$.} 
The Wilson loops appearing in Eq.~\eqref{GE} are computed on the paths
made up of the following quark $[+]$ - antiquark $[-]$ straight-line
paths (see Fig.~\ref{fig:1}),
\begin{equation}
{\cal C}^{\,(T)}_1 : {X}_{E1}^{\pm}(\tau) =  \pm
u_1 \tau + z
+ f^{\pm}_1 R_{1} , \quad
{\cal C}^{\,(T)}_2 : {X}_{E2}^{\pm}(\tau) =
\pm u_2 \tau 
+ f^{\pm}_2 R_{2},
\label{trajE}
\end{equation}
with $\tau\in [-T,T]$, and closed by straight-line paths in the
transverse plane at $\tau=\pm T$. The four-vectors $u_{1,2}$ are
chosen to be $u_{1,2}=(\pm\sin\frac{\theta}{2}, \vec{0}_{\perp}, 
\cos\frac{\theta}{2})$, $\theta$ being the angle formed by the two
trajectories, i.e., $u_1\cdot u_2 = \cos\theta$. Moreover,
$R_{i} = (0,\vec{R}_{i\perp},0)$, $z = (0,\vec{z}_{\perp},0)$ and
$f^+_i \equiv 1-f_i$, $f^{-}_i \equiv -f_i$. 
We define also the Euclidean and Minkowskian correlation functions
with the infrared cutoff removed as
\begin{equation}
  \label{eq:corem} 
  \begin{aligned}
\displaystyle {\cal C}_E(\theta;\vec{z}_\perp;\nu_1,\nu_2) &\equiv
\lim_{T\to\infty} {\cal G}_E(\theta;T;\vec{z}_\perp;\nu_1,\nu_2)\, ,\\ 
\displaystyle {\cal C}_M(\chi;\vec{z}_\perp;\nu_1,\nu_2) &\equiv
\lim_{T\to\infty} {\cal G}_M(\chi;T;\vec{z}_\perp;\nu_1,\nu_2)\, .
  \end{aligned}
\end{equation}
The dipole-dipole scattering amplitude is then obtained from
${\cal C}_E(\theta;\ldots)$, with $\theta\in(0,\pi)$, by means of
analytic continuation as 
\begin{equation}
  \label{eq:scatt-loop}
  \begin{aligned}
    {\cal M}_{(dd)} (s,t;\nu_1,\nu_2) 
    &\equiv -i~2s \displaystyle\int d^2 \vec{z}_\perp
    e^{i \vec{q}_\perp \cdot \vec{z}_\perp}
    {\cal C}_M(\chi \simeq \log(s/m^2); \vec{z}_\perp;\nu_1,\nu_2) 
    \\
    &  = -i~2s \displaystyle\int d^2 \vec{z}_\perp
    e^{i \vec{q}_\perp \cdot \vec{z}_\perp}
    {\cal C}_E(\theta\to -i\chi \simeq -i\log(s/m^2);
    \vec{z}_\perp;\nu_1,\nu_2)\, , 
  \end{aligned}
\end{equation}
with $\chi\in\mathbb{R}^+$, and where $s$ and $t = -|\vec{q}_\perp|^2$
($\vec{q}_\perp$ being the transferred momentum) are the usual
Mandelstam variables (for a detailed discussion on the analytic
continuation see Ref.~\cite{GM2009}, where we have shown, on
nonperturbative grounds, that the required analyticity hypotheses are
indeed satisfied). The restrictions on the domains of the variables
$\theta$ and $\chi$ cause no loss of generality, due to the symmetries
of the Euclidean and Minkowskian theories~\cite{crossing1,crossing2}. 

\begin{figure}[t]
  \centering
  \includegraphics[width=0.4\textwidth]{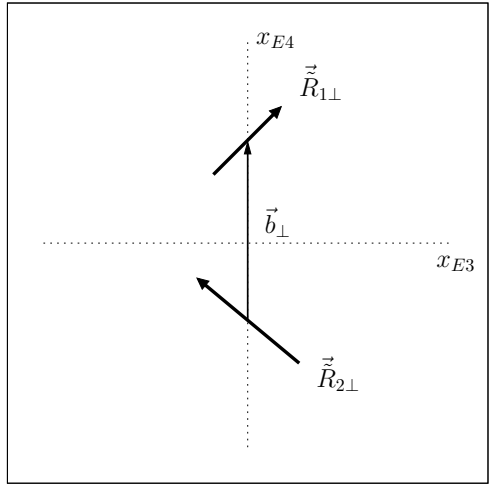}
  \caption{Relevant configuration of dipoles in Euclidean space after
    aligning the impact parameter with Euclidean ``time''.}
  \label{fig:2}
\end{figure}

For our purposes, it is convenient to exploit the rotation invariance
of the Minkowskian theory in order to fix the direction of $\vec
z_\perp$ along, say, the 2-axis. Indeed, dropping all the variables
that are irrelevant here, and choosing a rotation of angle $\varphi$
around the 1-axis in such a way that ${\cal R}_\varphi\vec{b}_\perp =\vec
z_\perp$,\footnote{Here ${\cal R}_\varphi$ denotes the restriction of the
  rotation to the $(2,3)$-plane.} with $\vec{b}_\perp = (b,0)$ and
$b=|\vec{z}_\perp|$, we have  
\begin{equation}
  \label{eq:rotations2}
  \begin{aligned}
    &\int d^2\vec{R}_{1\perp} \,|\psi_1(\vec{R}_{1\perp})|^2 \int
    d^2\vec{R}_{2\perp} \,
    |\psi_2(\vec{R}_{2\perp})|^2 \int d^2 \vec{z}_\perp\,
    e^{i \vec{q}_\perp \cdot \vec{z}_\perp}\,
    {\cal C}_M( \vec{z}_\perp;\vec{R}_{1\perp},\vec{R}_{2\perp}) 
\\ 
= &\int d^2 \vec{z}_\perp\, e^{i \vec{q}_\perp \cdot
  \vec{z}_\perp}\int d^2\vec{R}_{1\perp}\, |\psi_1({\cal 
  R}_\varphi\vec{R}_{1\perp})|^2 \int 
    d^2\vec{R}_{2\perp} \,
    |\psi_2({\cal R}_\varphi\vec{R}_{2\perp})|^2 \,
    {\cal C}_M(\vec b_\perp;\vec{R}_{1\perp},\vec{R}_{2\perp}) \,,
  \end{aligned}
\end{equation}
and we can write $\int d^2 \vec{z}_\perp = \int_0^\infty dbb
\int_0^{2\pi} d\varphi$. 
Expression Eq.~\eqref{eq:rotations2} simplifies in two cases. If
$\vec{q}_\perp=0$, the 
only dependence on the orientation of $\vec{z}_\perp$ appears in the
wave functions, so that we can treat the angular integration over
$\varphi$ as part of the averaging over the dipole variables, i.e., we
can write 
\begin{equation}
  \label{eq:rotations2_bis}
  \begin{aligned}
  {\cal M}_{(hh)}(s,0)  &=  -4\pi i s \int_0^{2\pi}\f{d\varphi}{2\pi}
\int d^2\vec{R}_{1\perp}\, |\psi_1({\cal 
  R}_\varphi\vec{R}_{1\perp})|^2 \int 
    d^2\vec{R}_{2\perp} \,
    |\psi_2({\cal R}_\varphi\vec{R}_{2\perp})|^2 \, \\
&\phantom{=}\times \int_0^\infty db\,b\, {\cal C}_M(\vec
b_\perp;\vec{R}_{1\perp},\vec{R}_{2\perp}) \equiv -4\pi i s\, \lla \int_0^\infty
db\,b\, {\cal C}_M(\vec 
b_\perp;\vec{R}_{1\perp},\vec{R}_{2\perp})  \rra_\varphi \,, 
  \end{aligned}
\end{equation}
where $\lla 1 \rra_\varphi = 1$. Notice that we are not making any
assumption on the wave functions. The other case is that in which the
wave functions are independent of the orientation of the dipoles: this is
the case, for example, if one considers amplitudes for unpolarised
scattering. Under this condition, ${\cal R}_\varphi$ drops from the
wave functions in Eq.~\eqref{eq:rotations2}, and after carrying out
the integration over $\varphi$ one obtains the following simple form
for the meson-meson scattering amplitude, 
\begin{equation}
  \label{eq:simple-scatt}
  {\cal M}_{(hh)}(s,t) = -4\pi i s \,\lla \int_0^\infty db b\,
  J_0(b\sqrt{-t})\,{\cal C}_M(\chi \simeq \log(s/m^2);
  \vec{b}_\perp;\nu_1,\nu_2) \rra_0 \,,
\end{equation}
where by the subscript 0 we indicate explicitly that the wave
functions are rotation-invariant. We note here that in most
phenomenological applications of the nonperturbative approach to
{\it soft} high energy scattering, the hadron wave functions are
chosen to be invariant under rotations and under the exchange $f_i\to
1-f_i$ (see Refs.~\cite{Dosch,LLCM1} and also~\cite{pomeron-book}, \S
8.6, and references therein).  

Clearly, ${\cal C}_M(\chi ;\vec{b}_\perp;\nu_1,\nu_2) = {\cal
  C}_E(\theta\to -i\chi;\vec{b}_\perp;\nu_1,\nu_2)$, due to the
analytic continuation relations. Furthermore, we can exploit the
$O(4)$ invariance of the Euclidean theory to show that 
\begin{equation}
  \label{eq:newcorrelator}
  {\cal C}_E(\theta;\vec{b}_\perp;\nu_1,\nu_2) = \tilde{\cal
    C}_E(\theta;b;\nu_1,\nu_2)  = \lim_{T\to\infty} \tilde{\cal
    G}_E(\theta;T;b;\nu_1,\nu_2)\,,
\end{equation}
where $\tilde{\cal G}_E$ is the correlation function of two Wilson
loops computed on new paths $\tilde{\cal C}^{\,(T)}_{1,2}$, 
\begin{equation}
  \label{eq:newcorr_2}
   \tilde{\cal G}_E(\theta;T;b;\nu_1,\nu_2)\equiv
\dfrac{\langle {\cal W}_E[\tilde{\cal C}^{\,(T)}_1]{\cal
    W}_E[\tilde{\cal C}^{\,(T)}_2]  
\rangle_E}{\langle {\cal W}_E[\tilde{\cal C}^{\,(T)}_1]  \rangle_E
\langle {\cal W}_E[\tilde{\cal C}^{\,(T)}_2] \rangle_E } - 1\,,
\end{equation}
with $\tilde{\cal C}^{\,(T)}_{1,2}$ obtained from ${\cal
  C}^{\,(T)}_{1,2}$ by rotating the transverse separation $(0,\vec
b_\perp,0)$ along the Euclidean ``time''  direction $x_{E4}$, see
Fig.~\ref{fig:2}. Explicitly, these paths are defined by 
\begin{equation}
  \label{eq:newpaths_newdef}
  \tilde{\cal C}^{\,(T)}_1: \tilde X_{E1}^{\pm}(\tau) =
\pm \tilde u_1\tau + \tilde z + f^{\pm}_1\tilde R_1
\,, \quad \tilde{\cal C}^{\,(T)}_2 : \tilde X_{E2}^{\pm}(\tau) = 
\pm \tilde u_2\tau  + f^{\pm}_2\tilde R_2\,,
\end{equation}
with $\tau\in[-T,T]$, and closed by appropriate straight-line paths at 
$\tau=\pm T$. Here
\begin{equation}
  \label{eq:newvect}
  \begin{aligned}
    \tilde u_{1,2} &=
    \left(\cos\f{\theta}{2},\pm\sin\f{\theta}{2},0,0\right)\,,  
       &&&
    \tilde z &=(0,0,0,b)\,, \\
    \tilde R_{i}&= 
     (0,0,\vec{\tilde R}_{i\perp})=(0,0,r_i\sin\phi_i ,r_i\cos\phi_i) \,,
     &&& &i=1,2\,, 
  \end{aligned}
\end{equation}
where $r_i=|\vec{R}_{i\perp}|$, $\phi_i$ is the angle formed by
$\vec{b}_\perp$ and $\vec{R}_{i\perp}$, and $f^\pm_i$ have been
defined after Eq.~\eqref{trajE}. For future utility, we define also
the paths $\tilde{\cal C}_0^{\,(T)}(\nu_i)$, 
\begin{equation}
  \label{eq:newpaths_zero}
  \tilde{\cal C}^{\,(T)}_0(\nu_i):
\tilde X_{E0}^{\pm}(\tau) =
\pm \tilde
    u_0\tau + f^{\pm}_i\tilde R_i
\,, \quad u_0=(1,0,0,0)\,,
\end{equation}
again with $\tau\in[-T,T]$, and closed by appropriate straight-line
paths. These are nothing but rectangular paths centered\footnote{Here
 by ``center'' of the loop we mean the ``center of mass'' of the
 dipole at $\tau=0$, i.e., $f_i  X_{E0}^+(0) + (1-f_i)X_{E0}^-(0)$.}
at the origin, and with the ``long'' side parallel to direction
1. Obviously, ${\cal C}_M(\chi;\vec{b}_\perp;\nu_1,\nu_2) =
\tilde{\cal C}_E(\theta\to -i\chi;b;\nu_1,\nu_2)$, so that
$\tilde{\cal C}_E$ encodes all the relevant information on the
scattering amplitude. 

We finally mention that the amplitude for meson-{\it anti}meson
scattering is obtained from Eq.~\eqref{eq:scatt-loop} by replacing
$\vec{R}_{2\perp},f_2 \to -\vec{R}_{2\perp},1-f_2$, or equivalently by
performing the analytic continuation $\chi \to i\pi -\chi$ of the
Minkowskian correlator ${\cal C}_M$, thanks to the crossing-symmetry
relations discussed in Refs.~\cite{crossing1,crossing2}, 
\begin{equation}
  \label{eq:cross_rel}
  {\cal C}_M(\chi; \vec{z}_\perp;\nu_1,\bar\nu_2)=
  {\cal C}_M(\chi; \vec{z}_\perp;\bar\nu_1,\nu_2)=
{\cal C}_M(i\pi-\chi; \vec{z}_\perp;\nu_1,\nu_2)\,,
\end{equation}
where $\bar\nu_i=(-\vec{R}_{i\perp},1-f_i)$. Notice that for hadronic
wave functions invariant under rotations and under the exchange
$f_i\to 1-f_i$ [see after Eq.~\eqref{eq:simple-scatt}], the scattering 
amplitude is automatically crossing-symmetric.  

\nsection{Relating hadronic total cross sections and the QCD spectrum: outline} 
\label{sec:outline}

As we have stated in the Introduction, the purpose of this paper is to
obtain new insights on the high-energy behaviour of hadronic total
cross sections, by relating the Wilson-loop correlation functions from
which the scattering amplitudes are built in the {\it soft} high
energy regime to the spectrum of QCD. As this involves a certain
number of rather technical steps, we want to provide first a general
outline of our argument, to make it easier for the reader to follow
the detailed discussion of the following Sections. 

The starting point is to re-express the relevant Euclidean correlation
function in the operator language. This requires the introduction of
the Euclidean Wilson loop {\it operator} $\hat\W_E$, which will be
defined precisely in Eq.~\eqref{eq:wl-opE} below. In terms of
$\hat\W_E$, Wilson-loop correlation functions in the functional-integral
formalism are rewritten as vacuum expectation values of $T$-ordered
products of Wilson loop operators. For our purposes, it is convenient
to work with the correlation function $\tilde{\cal G}_E$ defined in
Eq.~\eqref{eq:newcorr_2}, for which the separation along Euclidean
``time'' is equal to the impact-parameter distance $b$ in the
scattering process. For sufficiently large $b$, so that there are no
(Euclidean) time-ordering issues [see Eq.~\eqref{eq:condition}], the
relevant correlator reads (up to normalisation factors)
\begin{equation}
  \label{eq:corr_intro}
{\la {\W}_E[\tilde{\C}^{\,(T)}_1]{\W}_E[\tilde{\C}^{\,(T)}_2] 
\ra_E}
=
{\la 0 |
\hat{\W}_E[\tilde{\C}^{\,(T)}_1]\hat{\W}_E[\tilde{\C}^{\,(T)}_2]  
| 0 \ra}
\,,
\end{equation}
where the paths $\tilde{\C}^{\,(T)}_{1,2}$ have been defined above in
Eq.~\eqref{eq:newpaths_newdef}. 

The form Eq.~\eqref{eq:corr_intro} of the correlation function is
suitable for inserting a complete set of states between the two Wilson
loops. For this purpose we use asymptotic states characterised by
their particle content, and by the momentum and third component of the
spin of each particle. Denoting by $\alpha$ a generic state, and
exploiting the Euclidean symmetries, one finds
\begin{equation}
  \label{eq:corr_intro2}
  \begin{aligned}
{\la {\W}_E[\tilde{\C}^{\,(T)}_1]{\W}_E[\tilde{\C}^{\,(T)}_2] 
\ra_E}  
&= \sum_\alpha
{\la 0 |
\hat{\W}_E[\tilde{\C}^{\,(T)}_1]|\alpha\ra \la\alpha
|\hat{\W}_E[\tilde{\C}^{\,(T)}_2]   
| 0 \ra} 
\\ &= \sum_\alpha e^{-b E_\alpha}e^{i\theta S_3^{(\alpha)}}
{\la 0 |
\hat{\W}_E[\tilde{\C}^{\,(T)}_0(\nu_1)]|\alpha_{\f{\theta}{2}}\ra
\la\alpha_{-\f{\theta}{2}} 
|\hat{\W}_E[\tilde{\C}^{\,(T)}_0(\nu_2)]   
| 0 \ra}
\,,
  \end{aligned}
\end{equation}
where the sum over $\alpha$ includes also the appropriate phase-space
integration over the particles' momenta. Here $E_\alpha$ and
$S_3^{(\alpha)}$ are the total energy and total third component of the
spin for state $\alpha$, respectively, and the paths
$\tilde{\C}^{\,(T)}_0(\nu_{1,2})$, which have been defined in
Eq.~\eqref{eq:newpaths_zero}, are independent of $b$ and
$\theta$. Moreover, the states $|\alpha_{\pm\f{\theta}{2}}\ra$ are
obtained from $|\alpha\ra$ by performing a rotation of the momenta of
$\pm\f{\theta}{2}$ around the third axis [see
Eq.~\eqref{eq:more_not_bis}]. The use of time-translation invariance
in the second line of Eq.~\eqref{eq:corr_intro2} allows to completely
expose the dependence on $b$, while the use of rotation invariance
allows to shift the dependence on $\theta$ from the loops to the
momenta of the particles in the intermediate states; a further simple
change of variables allows to expose the $\theta$-dependence almost
entirely [see Eqs.~\eqref{eq:xpm}--\eqref{eq:more_not3}], yielding
\begin{equation}
  \label{eq:corr_intro2_bis}
  \begin{aligned}
\tilde{\cal C}_E
&= \lim_{T\to\infty} \dfrac{\la 0 |
  {\W}_E[\tilde{\C}^{\,(T)}_1]{\W}_E[\tilde{\C}^{\,(T)}_2]  
| 0 \ra} {\la 0 | \hat{\W}_E[\tilde{\C}^{\,(T)}_1] | 0 \ra
 \la 0 | \hat{\W}_E[\tilde{\C}^{\,(T)}_2]| 0 \ra } - 1
\\ &=  \sum_{\alpha\ne 0} \f{e^{-b E_\alpha}e^{i\theta
    S_3^{(\alpha)}}}{(\sin\theta)^{{\cal N}_\alpha}} 
M_\alpha(\theta;\nu_1,\nu_2) =\sum_{\alpha\ne 0} \delta C_\alpha
\,,
  \end{aligned}
\end{equation}
where ${\cal N}_\alpha$ is the number of particles in state $\alpha$,
$M_\alpha$ is given by the Wilson-loop matrix elements expressed in
terms of the new variables [times appropriate phase-space factors, see
Eq.~\eqref{eq:not_G_3}], and we have taken the physical limit
$T\to\infty$ [see Eqs.~\eqref{eq:newcorrelator} and
\eqref{eq:irremoved}].    

The next step is to perform the analytic continuation to Minkowski
space, which for the correlation function $\tilde{\cal C}_E$ reduces
to taking $\theta\to -i\chi$, and to study the large-$\chi$ limit. To
this extent, we make the crucial assumption that the analytic 
continuation can be carried out term by term, i.e., we assume that the
analytic continuation can be performed independently for each term
$\delta C_\alpha$ in the sum over states in
Eq.~\eqref{eq:corr_intro2_bis}. This requires that the sum has
``good'' convergence properties (e.g., uniform convergence in
$\theta$), so that analytic continuation and summation commute. Under
this assumption, it is easy to carry out the analytic continuation,
and to determine separately the leading energy dependence of each term
in the sum in the physical limit of large $\chi \sim \log s$. This is
due to the fact that, after analytic continuation, the function
$M_\alpha(\theta\to -i\chi;\nu_1,\nu_2)$ in
Eq.~\eqref{eq:corr_intro2_bis} becomes independent of $\chi$ at large 
$\chi$ [see Eqs.~\eqref{eq:not_G_3_an_con}--\eqref{eq:def_large_chi}]. 
Assuming that it is a finite nonzero quantity (see however footnote
\ref{foot:extrachi}), it is therefore possible to read off the leading
power of $s\sim e^\chi$ for each contribution directly from
Eq.~\eqref{eq:corr_intro2_bis}. One can easily see that at fixed
particle content the dominant contributions come from states with
maximal total spin. Furthermore, it is clear from 
Eq.~\eqref{eq:corr_intro2_bis} that at large $b$ each contribution
dies off exponentially. More precisely, for maximal total spin the
leading term in $\delta C_{\alpha}$ receives a factor $e^{\chi[s^{(a)}-1]}e^{-b
  m^{(a)}}$ from each particle, i.e., up to $\chi,b$-independent
factors and inverse powers of $b$ one finds
\begin{equation}
  \label{eq:corr_intro4}
  \delta C_\alpha \sim \prod_a \left[e^{\chi[s^{(a)}-1]}e^{-b
      m^{(a)}}\right]^{n_a(\alpha)}\,,
\end{equation}
with $m^{(a)}$ and $s^{(a)}$ respectively the mass and spin 
of particles of type $a$, and $n_a(\alpha)$ the corresponding
occupation number in state $\alpha$. In physical terms, this means
that states containing only particles of type $a$ contribute
appreciably to the correlator only up to impact-parameter distances of
the order of the ``effective radius'' $R_{\rm
  eff}^{(a)}=\chi[s^{(a)}-1]/m^{(a)}$ [or of an appropriate weighted
average of the effective radii, if different species of particles are
present, see Eq.~\eqref{eq:eff_rad}]. 

The final step consists in realising that, for our purposes, the
relevant contributions to the Wilson-loop correlator come from states
containing only a single type of particles, namely those with maximal
``effective radius''. This is because to obtain the elastic scattering
amplitude and the total cross section one has to integrate over the
impact parameter, and the dominant contributions to the integrals in
the large-$\chi$ limit come precisely from particles with maximal
``effective radius''. In turn, this implies that the leading relevant 
contributions to the correlator depend on $b$ only through the
combination $z=z(\chi,b)=e^{\chi(\tilde s-1)}e^{-b \tilde m}$, where
$\tilde m$ and $\tilde s$ are respectively the mass and spin of the
particle maximising the ratio $(s^{(a)}-1)/m^{(a)}$. More precisely,
up to constant factors, each of the $n$-particle sectors contributes a
term proportional to $w^n$, where $w=w(\chi,z(\chi,b)) \propto
z/\chi^\lambda$ for some real $\lambda$, whose precise value turns out
to be irrelevant for the leading asymptotic behaviour of the total
cross section.  

Summarising, under the analyticity and finiteness assumptions
mentioned above, it is possible to show that at large $\chi$ the
relevant (Minkowskian) Wilson-loop correlation function, ${\cal
  C}_M(\chi ;\vec{b}_\perp;\nu_1,\nu_2)$, depends only on a specific
combination, $w$, of $\chi$ and $b$, i.e.,  
\begin{equation}
  \label{eq:corr_intro5}
{\cal C}_M(\chi;\vec{b}_\perp;\nu_1,\nu_2)
\mathop\sim_{s\to\infty}g(w;\nu_1,\nu_2)-1\,,   
\end{equation}
[see Eq.~\eqref{eq:cmb}] for sufficiently large $b$ [see
Eq.~\eqref{eq:condition}]. Furthermore, one finds that the relevant
features of the detailed form of $w$ depend only on the spectrum of
the theory. This constitutes the first part of our program, and will
be discussed in Section \ref{sec:2}.  

Having derived the large-$\chi$ behaviour of the relevant Wilson-loop
correlation function, one can study the consequences for the 
asymptotic behaviour of hadronic total cross sections. An essential
ingredient here is the unitarity constraint, which provides bounds on
the scattering amplitude in impact-parameter space. As we argue in
Section \ref{sec:3}, the unitarity constraint translates into a bound
on the relevant Minkowskian correlator, i.e., $|{\cal C}_M(\chi
;\vec{b}_\perp;\nu_1,\nu_2)+1|\le 1$, which in turn implies that $g$
in Eq.~\eqref{eq:corr_intro5} is bounded [see
Eq.~\eqref{eq:unitarity_g}]. This immediately allows to 
identify the large-$b$ region as the one giving the dominant
contribution to the total cross sections, and to obtain a
``Froissart-like'' bound on the total cross sections [see
Eq.~\eqref{eq:J_split_bound}], 
\begin{equation}
  \label{eq:corr_intro6}
\sigma_{\rm tot}^{(hh)}(s)
 \mathop\lesssim_{s \to \infty} 
  4\pi\f{(\tilde s -1)^2}{\tilde m^2}\left(\log \f{s}{m^2}\right)^2\,,  
\end{equation}
where $\tilde s$ and $\tilde m$ have been defined above. If $g(w)$ is
either vanishing or oscillating at large $w$, one can derive a
stronger result, namely one can {\it predict} the asymptotic 
behaviour of $\sigma_{\rm tot}^{(hh)}$ and show that it is universal,
i.e., independent of the kind of hadrons involved. Explicitly, one
finds [see Eq.~\eqref{eq:sigma_universal}]
\begin{equation}
  \label{eq:corr_intro7}
\sigma_{\rm tot}^{(hh)}(s)
 \mathop\simeq_{s \to \infty} 
  2\pi\f{(\tilde s -1)^2}{\tilde m^2}\left(\log \f{s}{m^2}\right)^2\,.  
\end{equation}
Remarkably, in this case the prefactor of $\log^2 s$ is shown to be
entirely determined by the spectrum of QCD, and can be predicted by
finding the type of particle with maximal effective radius, as
explained above. 
The detailed discussion of these issues, and a few results on the
elastic scattering amplitudes, are reported in Section \ref{sec:3}.  

\nsection{Wilson-loop correlation function and the hadronic spectrum}
\label{sec:2}

In this Section we will show how the relevant Wilson loop correlator
can be related to the QCD spectrum, discussing in full detail the
first part of the argument outlined above in Section
\ref{sec:outline}. The consequences of our results for the hadronic
total cross sections and elastic scattering amplitudes will be
discussed in Section \ref{sec:3}. 

\subsection{Wilson loop in the operator formalism}

The ``good'' definition of the Wilson loop operator in Minkowski
space, preserving its gauge invariance, is the following~\cite{CT}:
\begin{equation}
  \label{eq:wl-op}
    \hat\W[\C] \equiv \f{1}{N_c}\, \Tr T\!P\!\exp\left\{-ig\oint_{\C} \hat
      A_\mu(x) dx^\mu 
  \right\}\,.
\end{equation}
Here and in the following the ``hat'' denotes an operator, and $TP$
stands for both time-ordering, acting on operators, and path-ordering,
acting on the colour matrices. Explicitly,
\begin{equation}
  \label{eq:TPord}
  \begin{aligned}
    TP&\left\{\hat A_\mu(x(\tau)) \hat A_\nu(x(\tau'))\right\}
    = \Big\{\Theta(x^0(\tau)-x^0(\tau'))\hat
      A^a_\mu(x(\tau))\hat A^b_\nu(x(\tau'))\\ & + 
      \Theta(x^0(\tau')-x^0(\tau))\hat A^b_\nu(x(\tau'))\hat
      A^a_\mu(x(\tau))\Big\}
\Big\{\Theta(\tau-\tau')t^at^b
      + \Theta(\tau'-\tau)t^bt^a\Big\}\,,
  \end{aligned}
\end{equation}
where $\Theta(x)$ is the Heaviside step function, and similarly in the
case of more terms. The bridge between the operator formalism and the
functional-integral formalism is provided by the relation
\begin{equation}
  \label{eq:wl_rel}
  \la \W[\C_1] \ldots \W[\C_n] \ra = \la 0 | {T}\left\{
\hat\W[\C_1] \ldots \hat\W[\C_n]
\right\} 
 | 0 \ra\,,
\end{equation}
where the time ordering is understood to act on the expansion of the
Wilson loops in products of field operators. The definition of the
Euclidean Wilson loop is the same as in Eq.~\eqref{eq:wl-op},
\begin{equation}
  \label{eq:wl-opE}
    \hat\W_E[\C] \equiv \f{1}{N_c}\, \Tr T\!P\!\exp\left\{-ig\oint_{\C} \hat
      A_{E\mu}(x_{E\mu}) dx_{E\mu} 
  \right\}\,,
\end{equation}
the only difference being that now the time-ordering is done with
respect to the Euclidean ``time''. Also,
\begin{equation}
  \label{eq:wl_relE}
  \la \W_E[\C_1] \ldots \W_E[\C_n] \ra_E = \la 0 | {T}\left\{
\hat\W_E[\C_1] \ldots \hat\W_E[\C_n]
\right\} 
 | 0 \ra\,,
\end{equation}
where $T$ is again time-ordering  with respect to the Euclidean
``time''. 

Let us now focus on the case of interest. Using
Eq.~\eqref{eq:wl_relE}, we can recast the correlation function
Eq.~\eqref{eq:newcorr_2} in terms of vacuum expectation values as
follows,  
\begin{equation}
  \label{eq:newcorr_op}
   \tilde{\cal G}_E(\theta;T;b;\nu_1,\nu_2) =
\dfrac{\la 0 |
  T\left\{\hat{\W}_E[\tilde{\C}^{\,(T)}_1]\hat{\W}_E[\tilde{\C}^{\,(T)}_2]
  \right\}    
| 0 \ra}{\la 0 | \hat{\W}_E[\tilde{\C}^{\,(T)}_1] | 0 \ra
\la 0 | \hat{\W}_E[\tilde{\C}^{\,(T)}_2]| 0 \ra } - 1\,.
\end{equation}
Since we are mainly interested in the large-distance behaviour of the
relevant Wilson-loop correlation function, we restrict our analysis to
the case of loops that do not overlap in Euclidean ``time'', 
which are characterised by 
\begin{equation}
  \label{eq:condition}
  b > b_0(\nu_1,\nu_2)\equiv r_1[f_1 -\Theta(-\cos\phi_1)]\cos\phi_1 -
  r_2[f_2 - \Theta(\cos\phi_2)]\cos\phi_2 \quad (\ge 0)\,,
\end{equation}
in terms of the transverse distance and of the sizes $r_i$ and
orientations $\phi_i$ of the dipoles [see Eq.~\eqref{eq:newvect}]. 
In this case we can drop the time-ordering symbol in the numerator,
obtaining 
\begin{equation}
  \label{eq:newcorr_op2}
   \tilde{\cal G}_E(\theta;T;b;\nu_1,\nu_2)=
\dfrac{\la 0 | \hat{\W}_E[\tilde{\C}^{\,(T)}_1]\hat{\W}_E[\tilde{\C}^{\,(T)}_2] 
| 0 \ra}{\la 0 | \hat{\W}_E[\tilde{\C}^{\,(T)}_1] | 0 \ra
\la 0 | \hat{\W}_E[\tilde{\C}^{\,(T)}_2]| 0 \ra } - 1\,.
\end{equation}
In the following we will always assume that Eq.~\eqref{eq:condition}
is satisfied, unless explicitly stated, so that
Eq.~\eqref{eq:newcorr_op2} holds.

\subsection{Inserting a complete set of states}

The stage is now set to insert a complete set of states between the
Wilson loop operators. According to the usual assumptions, such a
complete set of states is made of the asymptotic ({\it in} or {\it
  out}) states of the theory, containing any number of the particles
of the theory (including bound states)~\cite{Weinberg}. We choose {\it
  in} states for definiteness; the analysis is of course unchanged if
one uses {\it out} states instead.

A generic asymptotic {\it in} state is characterised by its particle
content, and by the momenta and the third component of the spins of
the particles. We will denote by
\begin{equation}
  \label{eq:state}
  | \alpha,\{ \vec p\}_\alpha,\{s_3\}_\alpha ~;~in\ra
\end{equation}
a state with particle content $\alpha$, where
$\alpha=\big(n_1,n_2,\ldots\big)$ is a string made up of the
occupation numbers $n_a=n_a(\alpha)$ of the various particle species
$a=1,2,\ldots$, characterised by their mass $m^{(a)}$ and spin
$s^{(a)}$, and moreover by their baryon number, electric charge,
``strangeness'', ``charm'', ``bottomness'' and ``topness''. From now
on, the latter quantum numbers will be indicated collectively as
``discrete charges''. Here $\{ \vec p\}_\alpha=\{
(p_1,p_2,p_3)\}_\alpha$ and $\{ s_3\}_\alpha$ denote respectively the
sets of all the momenta $\vec p^{\,(a)i}$ and all the third components
of the spin $s_3^{(a)i}$, where the index $a=1,2,\ldots$ runs on the
particle species and $i=1,2,\ldots,n_a(\alpha)$ on the particles of
the same species. As we are interested in real-world QCD, we will
consider the case $m^{(a)}> 0 \,\forall a$; the inclusion of massless
particles presents no particular difficulty, and will be briefly
discussed in Appendix \ref{app:1_1}. 

Let us now define the projector on the $n$-particle sector as follows: 
\begin{equation}
  \label{eq:projectors}
  |n \ra \la n| \equiv \f{1}{n!}
  \sum_{\alpha} \delta_{{\cal N}_\alpha,n}\,
  {\cal P}_\alpha
\sum_{\{s_3\}_\alpha}\int d\Omega_\alpha \,
| \alpha,\{ \vec p\}_\alpha, \{s_3\}_\alpha
~;~in\ra
\la \alpha,\{ \vec p\}_\alpha,\{s_3\}_\alpha ~;~in|\,,
\end{equation}
where the sum is over the strings $\alpha$ with ${\cal
  N}_\alpha\equiv\sum_a n_a(\alpha)$ equal to $n$, the factor 
\begin{equation}
  \label{eq:count_fact}
  {\cal P}_\alpha = \f{n!}{\prod_a n_a(\alpha)!}
\end{equation}
is due to Bose/Fermi symmetry (having factorised a $1/n!$ for
convenience), and we have denoted 
\begin{equation}
  \label{eq:not_sta}
  \begin{aligned}
    \sum_{\{s_3\}_\alpha} &= \prod_{a,\,n_a(\alpha)\ne 0}
    \prod_{i=1}^{n_a(\alpha)}
    \left\{\sum_{s_3^{(a)i}=-s^{(a)}}^{s^{(a)}}\right\}\,,\\  
\int d\Omega_\alpha &= \prod_{a,\,n_a(\alpha)\ne 0}
    \prod_{i=1}^{n_a(\alpha)}\left\{\int
    \f{d^3p^{(a)i}}{(2\pi)^32\varepsilon^{(a)i}}\right\}\,, \quad
  \varepsilon^{(a)i}=\sqrt{\left(m^{(a)}\right)^2+\left(\vec
      p^{\,(a)i}\right)^2} \,.  
  \end{aligned}
\end{equation}
We are using the standard relativistic normalisation for the
states.\footnote{For example, for one-particle states $\la \vec 
  p^{\,\prime}, s_3' | \vec
  p, s_3 \ra = 2\varepsilon(2\pi)^3\delta^{(3)}(\vec
  p^{\,\prime} - \vec p)\delta_{s_3 s_3'}$, for
  two-particle states $\la \vec
  p^{\,\prime}_1,p^{\,\prime}_2, s_{3\,1}',s_{3\,2}' | \vec
  p_1, \vec p_2, s_{3\,1},s_{3\,2} \ra =
  2\varepsilon_1 2\varepsilon_2(2\pi)^3\delta^{(3)}(\vec 
  p^{\,\prime}_1 - \vec p_1)(2\pi)^3\delta^{(3)}(\vec 
  p^{\,\prime}_2 - \vec
  p_2)\delta_{s_{3\,1}',s_{3\,1}}\delta_{s_{3\,2}',s_{3\,2}}+
  1\leftrightarrow 2$, and so on.} 
With this definition, the expansion of Eq.~\eqref{eq:newcorr_op2} over 
a complete set of states reads
\begin{equation}
  \label{eq:expansion}
 \tilde{\cal G}_E(\theta;T;b;\nu_1,\nu_2)= \sum_{n=1}^\infty  
\f{\la 0 | \hat{\W}_E[\tilde{\C}^{\,(T)}_1] | n\ra}{\la 0 |
  \hat{\W}_E[\tilde{\C}^{\,(T)}_1] | 0 \ra }\f{\la n|
  \hat{\W}_E[\tilde{\C}^{\,(T)}_2] | 0 \ra} {\la 0 |
  \hat{\W}_E[\tilde{\C}^{\,(T)}_2]| 0 \ra }  
= \sum_{n=1}^\infty \f{1}{n!}G_n(\theta;T;b;\nu_1,\nu_2)\,,
\end{equation}
where we have introduced the notation
\begin{equation}
  \label{eq:not_G}
  \begin{aligned}
    G_n(\theta;T;b;\nu_1,\nu_2) &\equiv
    \sum_{\alpha} \delta_{{\cal N}_\alpha,n}\,
    {\cal P}_\alpha
    \sum_{\{s_3\}_\alpha}\int d\Omega_\alpha\, \f{\la 0 |
      \hat{\W}_E[\tilde{\C}^{\,(T)}_1] |\alpha, \{ \vec p\}_\alpha,
      \{s_3\}_\alpha ~;~in\ra }{\la 0 | 
      \hat{\W}_E[\tilde{\C}^{\,(T)}_1] | 0 \ra } 
     \\
    &\phantom{\sum_{\alpha,\,{\cal N}_\alpha=n}oo
    {\cal P}_\alpha
    \sum_{\{s_3\}_\alpha}\int d\Omega_\alpha\, }
\times \f{ \la \alpha,\{ \vec p\}_\alpha,\{s_3\}_\alpha
    ~;~in|
      \hat{\W}_E[\tilde{\C}^{\,(T)}_2]| 0 \ra} {\la 0 |
      \hat{\W}_E[\tilde{\C}^{\,(T)}_2]| 0 \ra }  
    \,.
  \end{aligned}
\end{equation}
It is important to notice that, since the Wilson loop carries no
flavour, contributions to this sum come only from
states with vanishing ``discrete charges''.

We exploit now the invariance of the theory under translations along
Euclidean ``time'' and under rotations to write
\begin{equation}
  \label{eq:traslrot}
  \begin{aligned}
     \hat{\W}_E[\tilde{\C}^{\,(T)}_1]& =
 e^{\hat H b} e^{-i\hat J_3\f{\theta}{2}}
\hat{\W}_E[\tilde{\C}_0^{\,(T)}(\nu_1)]
e^{i\hat J_3\f{\theta}{2}}e^{-\hat H b}\,, \\
\hat{\W}_E[\tilde{\C}^{\,(T)}_2] &=
 e^{i\hat J_3\f{\theta}{2}}
\hat{\W}_E[\tilde{\C}_0^{\,(T)}(\nu_2)]
e^{-i\hat J_3\f{\theta}{2}}\,,
\end{aligned}
\end{equation}
where $\hat H$ and $\hat J_3$ are the Hamiltonian and the third
component of the angular momentum, i.e., we re-express the relevant
Wilson loops in terms of a rectangular loop with one (long) side
parallel to the 1-axis, and one (short) side in the (3,4)-plane
[recall the definition of $\tilde{\C}_0^{\,(T)}(\nu_i)$,
Eq.~\eqref{eq:newpaths_zero}]. We can now write
\begin{equation}
  \label{eq:not_G_2}
  \begin{aligned}
    G_n(\theta;T;b;\nu_1,\nu_2) &=\sum_{\alpha}\delta_{{\cal
        N}[\alpha],n}\, {\cal P}_\alpha
    \sum_{\{s_3\}_\alpha}e^{i \theta
      S_3^{(\alpha)}(\{s_3\}_\alpha)}\int d\Omega_\alpha\, 
    e^{-bE_\alpha(\{\vec 
  p\}_\alpha)}\\ &\phantom{\sum_{\alpha,\,{\cal N}_\alpha=n}
    }\times
  W_\alpha^{\,(T)}(\{{\cal R}_{\f{\theta}{2}} \vec p\}_\alpha,
    \{s_3\}_\alpha;\nu_1)
    \overline W_\alpha^{\,(T)}(\{
    {\cal R}_{-\f{\theta}{2}}\vec p\}_\alpha, 
    \{s_3\}_\alpha;\nu_2) 
\,.
  \end{aligned}
\end{equation}
Here we have introduced some new notation, which we now explain. The 
total energy $E_\alpha$ and the total third component of the spin
$S_3^{(\alpha)}$ are given by
\begin{equation}
  \label{eq:more_not}
  \begin{aligned}
    E_\alpha(\{\vec p\}_\alpha) &= \sum_{a,\,n_a(\alpha)\ne
      0}\sum_{i=1}^{n_a(\alpha)} 
\varepsilon^{(a)i}=
\sum_{a,\,n_a(\alpha)\ne
      0}\sum_{i=1}^{n_a(\alpha)} \sqrt{\left(m^{(a)}\right)^2+\left(\vec
      p^{\,(a)i}\right)^2} \,, \\
    S_3^{(\alpha)}(\{s_3\}_\alpha) &= \sum_{a,\,n_a(\alpha)\ne
      0}\sum_{i=1}^{n_a(\alpha)} s^{\,(a)i}_3\,.
  \end{aligned}
\end{equation}
As the baryon number (and so the fermion number) must be zero for a
state to contribute, the total spin $S_3^{(\alpha)}$ must be an
integer. The Wilson-loop matrix elements are denoted by 
\begin{equation}
  \label{eq:def_w_al}
  \begin{aligned}
      W_\alpha^{\,(T)}(\{ \vec p\}_\alpha,
    \{s_3\}_\alpha;\nu_i) &= 
\f{\la 0 |\hat{\W}_E[\tilde{\C}^{\,(T)}_0(\nu_i)]
    | \alpha,\{ \vec p\}_\alpha,
    \{s_3\}_\alpha 
    ~;~in\ra }{ \la 0 |
    \hat{\W}_E[\tilde{\C}^{\,(T)}_0(\nu_i)] | 0\ra}\,,
\\ 
\overline W_\alpha^{\,(T)}(\{
    \vec p\}_\alpha, 
    \{s_3\}_\alpha;\nu_i) &=\f{\la \alpha,\{ \vec
  p\}_\alpha,
  \{s_3\}_\alpha 
    ~;~in|\hat{\W}_E[\tilde{\C}^{\,(T)}_0(\nu_i)]
    | 0 \ra}{ \la 0
    |\hat{\W}_E[\tilde{\C}^{\,(T)}_0(\nu_i)] |
    0\ra}\,. 
  \end{aligned}
\end{equation}
In Eq.~\eqref{eq:not_G_2}, we have denoted the rotated three-momenta 
by $\{ {\cal R}_{\pm \f{\theta}{2}}\vec p\}_\alpha$, where
\begin{equation}
  \label{eq:more_not_bis}
  {\cal R}_\varphi = \left(
    \begin{array}{ccc}
      \phantom{-}\cos\varphi & \sin\varphi & 0\phantom{-}\\
      -\sin\varphi & \cos\varphi & 0\phantom{-}\\
      \phantom{-}0 & 0 & 1\phantom{-} 
    \end{array}\right)\,,
\end{equation}
and it is understood that the rotation is applied to the momenta of
all the particles. For our purposes, for $\theta\ne 0,\pi$, it is
convenient to re-express the rotated three-momenta in terms of the
variables 
\begin{equation}
  \label{eq:xpm}
  x^{\,(a)i}_\pm = \cos\f{\theta}{2}p^{\,(a)i}_1 \pm
    \sin\f{\theta}{2}p^{\,(a)i}_2\,.
\end{equation}
We have
\begin{equation}
  \label{eq:more_not2}
  \begin{aligned}
    \vec
    p^{\,(a)i} 
    &= \left(\f{x^{\,(a)i}_+ +
        x^{\,(a)i}_-}{2\cos\f{\theta}{2}},\f{x^{\,(a)i}_+ -
        x^{\,(a)i}_-}{2\sin\f{\theta}{2}},p_3\right)\,, \\
    {\cal R}_{\pm \f{\theta}{2}}\vec
    p^{\,(a)i} &= \left(x^{\,(a)i}_\pm,\pm \cot\theta x^{\,(a)i}_\pm \mp
      \f{1}{\sin\theta}x^{\,(a)i}_\mp,p_3\right)\,. 
  \end{aligned}
\end{equation}
We will therefore write
\begin{equation}
  \label{eq:more_not3}
  \begin{aligned}
    &W_\alpha^{\,(T)}(\{{\cal R}_{\f{\theta}{2}} \vec p\}_\alpha,
    \{s_3\}_\alpha;\nu_i) =  
    W_\alpha^{\,(T)}(\{\left(x_+,\cot\theta x_+ -
      \textstyle\f{1}{\sin\theta}x_-,p_3\right)\}_\alpha,
    \{s_3\}_\alpha;\nu_i)\,,\\
    &\overline W_\alpha^{\,(T)}(\{{\cal R}_{-\f{\theta}{2}} \vec p\}_\alpha,
    \{s_3\}_\alpha;\nu_i) = 
    \overline W_\alpha^{\,(T)}(\{\left(x_-,- \cot\theta x_- +
      \textstyle\f{1}{\sin\theta}x_+,p_3\right)\}_\alpha,
    \{s_3\}_\alpha;\nu_i)\,,
    \\
    &E_\alpha(\{\vec p\}_\alpha) =  E_\alpha(\{(\textstyle\f{x_+ +
        x_-}{2\cos\f{\theta}{2}},\textstyle\f{x_+ -
        x_-}{2\sin\f{\theta}{2}},p_3)\}_\alpha)=
\displaystyle \sum_{a,\,n_a(\alpha)\ne
      0}\sum_{i=1}^{n_a(\alpha)} \varepsilon^{(a)i}\,,
  \end{aligned}
\end{equation}
where in terms of the new variables
\begin{equation}
  \label{eq:newomega} 
\varepsilon^{(a)i} = \sqrt{\left(m^{(a)}\right)^2+  
          \left(\textstyle\f{x^{(a)i}_+ +
              x^{(a)i}_-}{2\cos\f{\theta}{2}}\right)^2 
+\left(\textstyle\f{x^{(a)i}_+ - x^{(a)i}_-}{2\sin\f{\theta}{2}}\right)^2 +
\left(p^{(a)i}_3\right)^2}\,.
\end{equation}
Furthermore, it is easy to obtain the Jacobian for the change of 
variables, 
\begin{equation}
  \label{eq:more_not5}
  d^3p^{(a)i} = \f{1}{|\sin\theta|}\,dx_+^{(a)i} dx_-^{(a)i}
  dp_3^{(a)i}\,, 
\end{equation}
so that we can write
\begin{equation}
  \label{eq:more_not5bis}
  \begin{aligned}
    d\Omega_\alpha&=
    \f{1}{|\sin\theta|^{{\cal N}_\alpha}}
 dX^+_{\alpha}dX^-_{\alpha}
    d\tilde\Omega_\alpha 
    h_\alpha(\{(\textstyle\f{x_+ + 
        x_-}{2\cos\f{\theta}{2}},\textstyle\f{x_+ -
        x_-}{2\sin\f{\theta}{2}},p_3)\}) \,,\\ 
dX^\pm_{\alpha}
 &= \prod_{a,\,n_a(\alpha)\ne 0}
    \prod_{i=1}^{n_a(\alpha)}\f{dx^{(a)i}_\pm}{2\pi}\,, \quad
d\tilde\Omega _\alpha= \prod_{a,\,n_a(\alpha)\ne 0}
    \prod_{i=1}^{n_a(\alpha)}
    \f{dp^{(a)i}_3}{(2\pi)2\tilde\varepsilon^{(a)i}}\,, \\ 
h_\alpha&      = \prod_{a,\,n_a(\alpha)\ne 0}
    \prod_{i=1}^{n_a(\alpha)}
    \f{\tilde\varepsilon^{(a)i}}{\varepsilon^{(a)i}}\,,
\quad
    \tilde\varepsilon^{(a)i}=\sqrt{\left(m^{(a)}\right)^2+
      \left(p^{(a)i}_3\right)^2}   \,.
    \\ 
  \end{aligned}
\end{equation}
Restricting to $\theta\in(0,\pi)$, so that we can drop the absolute
value from the Jacobian, we can finally write 
\begin{equation}
  \label{eq:not_G_3}
  \begin{aligned}
G_n(\theta;T;b;\nu_1,\nu_2)    &=\f{1}{(\sin\theta)^n}\sum_{\alpha}
\delta_{{\cal N}_\alpha,n} \,
    {\cal P}_\alpha
    \sum_{\{s_3\}_\alpha}e^{i \theta S_3^{(\alpha)}(\{s_3\}_\alpha)}\int
     dX^+_{\alpha}\int dX^-_{\alpha}\int
    d\tilde\Omega_\alpha\,
\\    &\phantom{\sum_{\alpha,\,{\cal N}_\alpha=n}
     }\times
    h_\alpha(\{(\textstyle\f{x_+ + 
        x_-}{2\cos\f{\theta}{2}},\textstyle\f{x_+ -
        x_-}{2\sin\f{\theta}{2}},p_3)\}) e^{-b
      E_\alpha\left(\{(\f{x_+ + 
        x_-}{2\cos\f{\theta}{2}},\f{x_+ -
        x_-}{2\sin\f{\theta}{2}},p_3)\}_\alpha\right)}\\
    &\phantom{\sum_{\alpha,\,{\cal N}[\alpha]=n} 
      }\times
W_\alpha^{\,(T)}(\{\left(x_+,\cot\theta x_+ -
      \textstyle\f{1}{\sin\theta}x_-,p_3\right)\}_\alpha,
    \{s_3\}_\alpha;\nu_1)
\\ &\phantom{\sum_{\alpha,\,{\cal N}[\alpha]=n}
        }\times
\overline W_\alpha^{\,(T)}(\{\left(x_-,- \cot\theta x_- +
      \textstyle\f{1}{\sin\theta}x_+,p_3\right)\}_\alpha,
    \{s_3\}_\alpha;\nu_2)
\,.
  \end{aligned}
\end{equation}
Let us introduce one last piece of notation. Since we are interested
in the limit of infinite loop length, and we expect such a limit to
exist for all the matrix elements $W_\alpha^{(T)}$, $\overline
W_\alpha^{(T)}$ separately,\footnote{\label{foot:LSZ} This can be
  understood in the LSZ framework~\cite{LSZ1,LSZ2}, where the matrix
  elements get replaced by the vacuum expectation values of products
  of appropriate interpolating fields and the Wilson loop. Due to the
  short-range nature of strong interactions, those parts of the loop
  that are too distant from the interpolating fields do not interact
  with them, and give contributions only to the self-interaction of
  the loop. As these contributions get cancelled by the normalisation
  factor, one expects $W_\alpha^{(T)}$ and $\overline W_\alpha^{(T)}$
  to become almost constant beyond some ``critical'' loop length.} we
define  
\begin{equation}
  \label{eq:inf_T}
  C_n\equiv\lim_{T\to\infty}G_n\,, \quad
  W_\alpha\equiv\lim_{T\to\infty}W_\alpha^{(T)}\,, \quad
  \overline W_\alpha\equiv\lim_{T\to\infty}\overline W_\alpha^{(T)}\,,
\end{equation}
and so we write
\begin{equation}
  \label{eq:irremoved}
  \tilde{\cal C}_E = \sum_{n=1}^\infty \f{1}{n!}\, C_n\,.
\end{equation}

\subsection{Analytic continuation to Minkowski space}
\label{sec:an_cont} 

The expression Eq.~\eqref{eq:not_G_3}, in the limit $T\to\infty$, is
the starting point for the analytic continuation back to Minkowski
space, that we now discuss. At this point we make two crucial
analyticity assumptions:
\begin{enumerate}
\item the analytic
continuation can be performed term by term, i.e., separately for 
the contribution of each state; 
\item the matrix elements $W_\alpha$ and $\overline W_\alpha$,
  expressed in terms of the variables $x^{(a)i}_\pm$, are analytic in
  $\theta$, in a complex domain including the real segment $(0,\pi)$
  and the negative imaginary axis.\footnote{Technically, this amounts
    to asking for analyticity of $W_\alpha$ and $\overline W_\alpha$ 
   in the second component of the three-momenta of the particles.
   Notice that analyticity in all the first components of the momenta
   is {\it not} satisfied, as the translational invariance along
   direction 1 in the limit $T\to\infty$ leads to the appearence of a
   delta function imposing that the first component of the total
   momentum vanish.}
\end{enumerate}
The first assumption is especially strong, as it requires appropriate
convergence properties of the double series defined by
Eqs.~\eqref{eq:expansion} and \eqref{eq:not_G_3} (at least in the
limit $T\to\infty$). We will discuss later possible ways of partially
relaxing this condition.

Let us now consider the various terms of Eq.~\eqref{eq:not_G_3}, in
the limit $T\to\infty$, and analytically continuing $\theta$ in the
complex plane, i.e., replacing $\theta\to
\theta-i\chi$ with $\theta\in(0,\pi)$ and $\chi\in\mathbb{R}^+$. The
physical, Minkowskian quantity is obtained in the limit $\theta\to
0$.\footnote{This limit gives the physical amplitude in the direct
  channel. The limit $\theta\to \pi$ (with negative $\chi$) provides
  the amplitude in the crossed channel, see the discussion at the end
  of Section \ref{sec:1}.} Let us start from the total energy  
$E_\alpha=\sum_{a,i}\varepsilon^{(a)i}$. We have 
\begin{equation}
  \label{eq:an_con_E}
  \begin{aligned}
  \varepsilon^{(a)i} &\mathop\to_{\theta\to \theta-i\chi}
  \sqrt{\left(m^{(a)}\right)^2+   
          \left(\textstyle\f{x^{(a)i}_+ +
              x^{(a)i}_-}{2\cos\f{\theta-i\chi}{2}}\right)^2 
+\left(\textstyle\f{x^{(a)i}_+ - x^{(a)i}_-}{2\sin\f{\theta-i\chi}{2}}\right)^2 +
\left(p^{(a)i}_3\right)^2}\\
&= \sqrt{\left(m^{(a)}\right)^2+\left(p^{(a)i}_3\right)^2 + Q^{(a)i}}
= \sqrt{V^{(a)i}}\,, 
  \end{aligned}
\end{equation}
with
\begin{equation}
  \label{eq:an_con_E_2}
  \begin{aligned}
  Q^{(a)i} =&~ (A^{(a)i})^2\left(\cos\textstyle\f{\theta}{2}\right)^2
  + (B^{(a)i})^2\left(\sin\textstyle\f{\theta}{2}\right)^2 +
  [(A^{(a)i})^2-(B^{(a)i})^2](\cos\theta)^2 
  (\sinh\textstyle\f{\chi}{2})^2 \\ &+
  \f{i}{2}[(B^{(a)i})^2-(A^{(a)i})^2]\sin\theta\sinh\chi\,,\\ 
  A^{(a)i} =&~
 \f{1}{2} \f{x_+^{(a)i}+x_-^{(a)i}}{(\cos\f{\theta}{2}\cosh\f{\chi}{2})^2+
   (\sin\f{\theta}{2}\sinh\f{\chi}{2})^2}  \,,\\
  B^{(a)i} =&~
 \f{1}{2} \f{x_+^{(a)i}-x_-^{(a)i}}{(\sin\f{\theta}{2}\cosh\f{\chi}{2})^2+
   (\cos\f{\theta}{2}\sinh\f{\chi}{2})^2}  \,.
  \end{aligned}
\end{equation}
Convergence problems in the integration over $x_\pm^{(a)i}$ and
$p_3^{(a)i}$ may arise if there were regions with $\Re \varepsilon^{(a)i}
<0$. This can happen only if the phase of the argument of the square
root in Eq.~\eqref{eq:an_con_E},
$V^{(a)i}=|V^{(a)i}|e^{i\varphi^{(a)i}}$, grows beyond $\pm \pi$. In
turn, this can happen only if $V^{(a)i}$ crosses the negative real
axis, i.e., if there is a point where $\Im V^{(a)i} =0$ with $\Re
V^{(a)i} <0$. However, for $\theta\in (0,\pi)$, $\Im V^{(a)i} =0$
implies $\Im Q^{(a)i} =0$ and therefore $A^{(a)i}=\pm B^{(a)i}$, so that
$\Re Q^{(a)i} \ge 0$ and thus $\Re V^{(a)i} > 0$. As a consequence, as
long as $\theta\ne 0,\pi$, one has $\varphi^{(a)i}\in(-\pi,\pi)$,
which finally implies $\Re \varepsilon^{(a)i} > 0$ $\forall a,i$,
i.e., $\Re E_\alpha > 0$.\footnote{For
  a massless particle $m^{(a_0)}=0$ one has $\Re V^{(a_0)i} > 0$,
  except at $x_+^{(a)i}=x_-^{(a)i}=p^{(a)i}_3=0$ where $\Re
  V^{(a_0)i} = 0$. As a consequence, if there are massless particles
  in the spectrum one still has $\Re E_\alpha > 0$, except on 
  the set of states containing only massless particles
  of zero momentum, where $\Re E_\alpha =
  0$, but which is a set of zero measure.}  On the other hand, when
$\theta=0,\pi$ one has $\Im V^{(a)i} =0$ 
independently of the integration variables, while $\Re V^{(a)i}$ can
be negative. Therefore, one should keep in mind that the limits 
$\theta\to 0,\pi$ can be taken only after performing the integration:
a small but nonzero $\theta$ serves as a regularisation, that will be 
understood in the following. 

Notice also that, as long as $\theta\ne 0,\pi$, one has $V^{(a)i}\ne 0$, so
that no singularity appears in the quantity $h_\alpha$. Moreover, for 
$\theta=0,\pi$, these singularities are integrable, so that they cause 
no problem to the integration even in the limit $\theta\to 0,\pi$.

Having assumed analyticity of the matrix elements, there is no further 
problem in carrying out the analytic continuation to Minkowski space
(i.e., in taking $\theta\to 0$), obtaining 
\begin{equation}
  \label{eq:not_G_3_an_con}
  \begin{aligned}
C_n(-i\chi;b;\nu_1,\nu_2)    &=\left(\f{i}{\sinh\chi}\right)^n\sum_{\alpha}
\delta_{{\cal N}_\alpha,n} \,
    {\cal P}_\alpha
    \sum_{\{s_3\}_\alpha}e^{\chi S_3^{(\alpha)}(\{s_3\}_\alpha)}\\
&\phantom{\sum_{\alpha}
     }\times\int
     dX^+_{\alpha}\int dX^-_{\alpha}\int
    d\tilde\Omega_\alpha\,
    h_\alpha(\{(\textstyle\f{x_+ + 
        x_-}{2\cosh\f{\chi}{2}},\textstyle\f{i(x_+ -
        x_-)}{2\sinh\f{\chi}{2}},p_3)\}_\alpha)\\
    &\phantom{\sum_{\alpha}
      }\times e^{-b
      E_\alpha\left(\{(\f{x_+ + 
        x_-}{2\cosh\f{\chi}{2}},\f{i(x_+ -
        x_-)}{2\sinh\f{\chi}{2}},p_3)\}_\alpha\right)}\\
    &\phantom{\sum_{\alpha}
      }\times
W_\alpha(\{\left(x_+,i\coth\chi x_+ -
      \textstyle\f{i}{\sinh\chi}x_-,p_3\right)\}_\alpha,
    \{s_3\}_\alpha;\nu_1)
\\ &\phantom{\sum_{\alpha}
        }\times
\overline W_\alpha(\{\left(x_-,- i\coth\chi x_- +
      \textstyle\f{i}{\sinh\chi}x_+,p_3\right)\}_\alpha,
    \{s_3\}_\alpha;\nu_2)
\,,
  \end{aligned}
\end{equation}
where, as we have explained above, the expression for the energy
$E_\alpha$ is properly regularised. 

As we have explained in Section \ref{sec:1}, the Wilson-loop correlation
function encodes the scattering amplitude in the high-energy
regime. Therefore, physical results are obtained in the limit 
$\chi\to\infty$, that we now discuss. Since\footnote{Notice that we
  are taking $\chi\to\infty$ before (actually, without) taking
  $\theta\to 0$: we are assuming that the limit $\chi\to\infty$
  commutes with the integration (for $\theta\ne 0,\pi$).} 
\begin{equation}
  \label{eq:an_con_E_3}
  \begin{aligned}
    \varepsilon^{(a)i} &\mathop\to_{\theta\to \theta-i\chi}
    \sqrt{\left(m^{(a)}\right)^2+   
          \left(\textstyle\f{x^{(a)i}_+ +
              x^{(a)i}_-}{2\cos\f{\theta-i\chi}{2}}\right)^2 
+\left(\textstyle\f{x^{(a)i}_+ -
    x^{(a)i}_-}{2\sin\f{\theta-i\chi}{2}}\right)^2 + 
\left(p^{(a)i}_3\right)^2} \\
& \mathop\to_{\chi\to \infty} \sqrt{\left(m^{(a)}\right)^2 
          +\left(p^{(a)i}_3\right)^2} +{\cal O}(e^{-\chi}) =
        \tilde\varepsilon^{(a)i}+{\cal O}(e^{-\chi}) \,,
  \end{aligned}
\end{equation}
we have that $h_\alpha \to 1$ after analytic continuation and in the
large-$\chi$ limit, and that 
\begin{equation}
  \label{eq:an_con_E_4}
  E_\alpha\left(\{\textstyle (\f{x_+ + 
        x_-}{2\cosh\f{\chi}{2}},\f{i(x_+ -
        x_-)}{2\sinh\f{\chi}{2}},p_3)\}_\alpha\right)
    \mathop\to_{\chi\to \infty}  \tilde E_\alpha(\{p_3\}_\alpha) = 
    \sum_{a,\,n_a(\alpha)\ne 
      0}\sum_{i=1}^{n_a(\alpha)} \tilde\varepsilon^{(a)i} \,.
\end{equation}
Also, to leading order, $W_\alpha$ is independent of $x_-^{(a)i}$, and
$\overline W_\alpha$ is independent of $x_+^{(a)i}$. Finally, due to the
exponential prefactor $e^{\chi S^{(\alpha)}_3}$, for a given particle
content $\alpha$ the leading contribution comes from the spin
configuration in which $s_3^{(a)i} = s^{(a)}$ $\forall
i\in\{1,\ldots,n_a(\alpha)\}$, which we will denote as
$\{s_3=s\}_\alpha$. To leading order in $\chi$ we have therefore 
\begin{equation}
  \label{eq:not_G_3_an_con_large_chi_2}
  \begin{aligned}
C_n(-i\chi;b;\nu_1,\nu_2)    &\mathop \sim_{\chi\to\infty}(2i)^n\sum_{\alpha}
\delta_{{\cal N}_\alpha,n} \,
    {\cal P}_\alpha\, e^{\chi [S_3^{(\alpha)}(\{s_3=s\}_\alpha)-n]}\int
    d\tilde\Omega_\alpha\, e^{-b
      \tilde E_\alpha\left(\{p_3\}_\alpha\right)}\\
    &\phantom{\sum_{\alpha}
      }\times
 {\cal F}_\alpha(\{p_3\}_\alpha,\nu_1)
\overline{\cal F}_\alpha(\{p_3\}_\alpha,\nu_2)
\,,
  \end{aligned}
\end{equation}
where
\begin{equation}
  \label{eq:def_large_chi}
  \begin{aligned}
 {\cal F}_\alpha(\{p_3\}_\alpha;\nu_1) &\equiv \int 
 dX^+_{\alpha}\,
W_\alpha(\{\left(x_+,ix_+,p_3\right)\}_\alpha,
    \{s_3=s\}_\alpha;\nu_1)\,,\\
\overline{\cal F}_\alpha(\{p_3\}_\alpha;\nu_2)  &\equiv  \int
 dX^-_{\alpha}\,
\overline W_\alpha(\{\left(x_-,- i 
  x_-,p_3\right)\}_\alpha, 
    \{s_3=s\}_\alpha;\nu_2)
\,,
  \end{aligned}
\end{equation}
with corrections being of relative order ${\cal
  O}(e^{-\chi})$.\footnote{\label{foot:extrachi} Here we are assuming
  that ${\cal F}_\alpha$ and $\overline{\cal F}_\alpha$ are finite
  quantities, but it is of course possible that they are zero or
  infinite. In these cases, in
  Eq.~\eqref{eq:not_G_3_an_con_large_chi_2} they would be replaced by
  a finite quantity times a $\chi$-dependent suppression or
  enhancement factor, respectively. This would change quantitatively  
  the result obtained with our method, but not the qualitative features
  of our argument.}

The result above depends crucially on our analyticity assumptions,
which can however be relaxed. A possibility which is worth discussing
is that the term-by-term analytic continuation can be performed only
in some limited range of $\chi$ at any fixed value of $b$. Since a
larger $b$ makes the coefficient of $e^{\chi
  [S_3^{(\alpha)}(\{s_3=s\}_\alpha)-n]}$ smaller, in this case we
expect the range of $\chi$ to widen at larger impact parameter,
including higher and higher values of the energy. Turning the argument
around, we expect in this case that increasing the energy requires to
go to larger impact parameter to perform the term-by-term analytic
continuation. As we will discuss in the following, this could be
enough for our approach to work. 

\subsection{Large-$b$ behaviour}
\label{subsec:largeb}

Before discussing the physical consequences of our result for $C_n$,
Eq.~\eqref{eq:not_G_3_an_con_large_chi_2}, it is useful to determine
its behaviour for large impact parameter $b$. In order to do so, let
us perform the change of variables 
\begin{equation}
  \label{eq:large_b1}
  p^{(a)i}_3 = \f{\tilde p_3^{(a)i}}{\sqrt{b m^{(a)}}}\,.
\end{equation}
The integration measure becomes
\begin{equation}
  \label{eq:large_b2}
  \begin{aligned}
    d\tilde\Omega _\alpha&= \prod_{a,\,n_a(\alpha)\ne 0}
    \prod_{i=1}^{n_a(\alpha)}
    \f{d\tilde p_3^{(a)i}}{(2\pi)2\sqrt{b m^{(a)}}
      \sqrt{\left(m^{(a)}\right)^2+\left(\f{\tilde p_3^{(a)i}}{\sqrt{b
              m^{(a)}}}\right)^2}} 
    \\ &= 
\prod_{a,\,n_a(\alpha)\ne 0}
    \prod_{i=1}^{n_a(\alpha)} \f{1}{2\sqrt{bm^{(a)}}}
    \f{d\tilde p_3^{(a)i}}{(2\pi)m^{(a)}
      }\left(1+{\cal
        O}\left(\f{1}{b m^{(a)}}\left(\f{\tilde
            p_3^{(a)i}}{m^{(a)}}\right)^2\right)\right)  
    \,,
  \end{aligned}  
\end{equation}
while expanding the energy $\tilde E_\alpha$ in inverse powers of $b
m^{(a)}$ we obtain
\begin{equation}
  \label{eq:large_b3}
  \begin{aligned}
    \tilde E_\alpha(\{\textstyle\f{\tilde p_3}{\sqrt{b m}}\}_\alpha) =
   \displaystyle\sum_{a,\,n_a(\alpha)\ne 
      0}\sum_{i=1}^{n_a(\alpha)}\left[ m^{(a)} +
      \f{1}{2b}\left(\f{\tilde p_3^{(a)i}}{m^{(a)}}\right)^2 + {\cal
        O}\left(\f{m^{(a)}}{(b m^{(a)})^2}
        \left(\f{\tilde p_3^{(a)i}}{m^{(a)}}\right)^4\right)\right]\,.  
  \end{aligned}
\end{equation}
Assuming now that ${\cal F}_\alpha(\{0\}_\alpha;\nu_1)$
and  $\overline{\cal F}_\alpha(\{0\}_\alpha;\nu_2)$ are 
nonzero, where $\{0\}_\alpha$ denotes $p_3^{(a)i}=0~\forall\,a,i$, and
carrying out the integrations over $\tilde p_3^{(a)i}$, we obtain  
\begin{equation}
  \label{eq:not_G_3_an_con_large_chi_large_b_2}
  \begin{aligned}
C_n(-i\chi;b;\nu_1,\nu_2) & \mathop\sim_{\chi\to\infty,\,b\to\infty}
i^n\sum_{\alpha} \delta_{{\cal N}_\alpha,n} \,
    {\cal P}_\alpha
{\cal F}_\alpha(\{0\}_\alpha;\nu_1)
    \overline{\cal F}_\alpha(\{0\}_\alpha;\nu_2)
    \\
    &\phantom{iiiooo}\times 
\prod_{a}\left( \f{1}{\sqrt{2\pi b m^{(a)}}}
e^{\chi[s^{(a)}-1]}e^{-b m^{(a)}}\right)^{n_a(\alpha)}\,,
  \end{aligned}
\end{equation}
with corrections being of relative order ${\cal O}(e^{-\chi})$ and
${\cal O}(b^{-1})$. The finiteness assumption is not crucial: if 
${\cal F}_\alpha(\{0\}_\alpha;\nu_1)$ and/or  $\overline{\cal
  F}_\alpha(\{0\}_\alpha;\nu_2)$ vanish, extra inverse powers of $b$
appear, which will not affect dramatically the high energy behaviour
of the amplitude. A detailed discussion of this issue is provided in
Appendix \ref{app:1_1}, where the effects due to the presence of massless
particles in the spectrum are also considered.

The physical interpretation of
Eq.~\eqref{eq:not_G_3_an_con_large_chi_large_b_2} is that the
contribution to $C_n$ of the states $\alpha$, characterised by a given
particle content, is non-negligible as long as the impact-parameter
distance is smaller than or of the order of a critical ``effective
radius'',
\begin{equation}
  \label{eq:eff_rad}
  R_{\rm eff}^{[\alpha]}(s) \equiv \f{\sum_a
    n_a(\alpha)[s^{(a)}-1]}{\sum_a n_a(\alpha) m^{(a)}}\chi  
  = \f{\sum_a n_a(\alpha)m^{(a)} R^{(a)}_{\rm eff}(s)
  } {\sum_a n_a(\alpha) m^{(a)}}\,,
\end{equation}
growing like $\sim\!\log s$, but with a prefactor that depends on the
particle content. This means that while the ratio of effective radii
corresponding to different particle contents is constant, their
difference can grow logarithmically with energy. In the last passage
of Eq.~\eqref{eq:eff_rad} we have made explicit that the effective
radius for state $\alpha$ is the weighted average of the
single-particle effective radii, 
\begin{equation}
  \label{eq:eff_rad2}
  R^{(a)}_{\rm eff}(s)\equiv \f{s^{(a)}-1}{m^{(a)}} \chi \equiv
  l_0^{(a)}\chi\,. 
\end{equation}

\subsection{Large-$\chi$ behaviour}
\label{subsec:largechi}

It is clear from Eq.~\eqref{eq:not_G_3_an_con_large_chi_large_b_2}
that one cannot straightforwardly take the limit $\chi\to\infty$ of
the quantity $C_n$. Nevertheless, since we are ultimately 
interested in integrating over the impact parameter $b$ to determine
the elastic scattering amplitude and the total cross section, it would
be enough for our purposes if we could define a variable, which is a
suitable combination of $\chi$ and $b$, that encodes the energy and
impact-parameter dependencies in the high-energy limit. To this
extent, we define the quantity 
\begin{equation}
  \label{eq:change_var_0}
  z(\chi,b) \equiv e^{c\chi}e^{-M b}\,, 
\end{equation}
where the parameters $c$ and $M$ will be determined later, and we
re-express $C_n$ in terms of $z$ and $\chi$. Using the large-$\chi$,
large-$b$ expression
Eq.~\eqref{eq:not_G_3_an_con_large_chi_large_b_2}, 
and including explicitly the subleading terms,  we find
\begin{equation}
  \label{eq:not_G_3_an_con_large_chi_large_b_3}
  \begin{aligned}
C_n(-i\chi;b;\nu_1,\nu_2) 
&= i^n\sum_{\alpha}
\delta_{{\cal N}_\alpha,n} \,
    {\cal P}_\alpha
{\cal F}_\alpha(\{0\}_\alpha;\nu_1)
    \overline{\cal F}_\alpha(\{0\}_\alpha;\nu_2)
    \\
    &\phantom{iiiooo}\times 
\prod_{a}\left\{
  \f{e^{\chi[s^{(a)}-1-c\f{m^{(a)}}{M}]}z^{\f{m^{(a)}}{M}}}{\left[2\pi 
      \log\left(\f{e^{c\chi}}{z}\right)\f{m^{(a)}}{M}\right]^{\f{1}{2}}}
\right\}^{n_a(\alpha)}
\\ &\phantom{iiiooo}\times 
\left(1 + {\cal O}\left(\f{1}{ \log\f{e^{c\chi}}{z}}\right) +
   {\cal O}(e^{-\chi})
\right) 
\,,
  \end{aligned}
\end{equation}
where for clarity we have suppressed the dependence of $z$ on $\chi$
and $b$. We are considering here the case of only massive particles in
the spectrum. If we now choose
\begin{equation}
  \label{eq:change_var_2}
  \f{c}{M} = \max_a \f{s^{(a)}-1}{m^{(a)}} = \max_a l_0^{(a)}\,,
\end{equation}
assuming that it exists and that it is positive,\footnote{If the
  maximum in Eq.~\eqref{eq:change_var_2} exists but is negative 
  or zero, we can take straightforwardly $\chi\to\infty$ in
  Eq.~\eqref{eq:not_G_3_an_con_large_chi_large_b_2}, obtaining either
  zero or a function of $b$ only. In turn, this leads to a vanishing or
  constant forward elastic scattering amplitude (and thus total cross section)
  at high energy.} we immediately see that
\begin{equation}
  \label{eq:change_var_3}
  \lim_{\chi\to\infty} e^{\chi[s^{(a)}-1-c\f{m^{(a)}}{M}]} = \left\{
    \begin{aligned}
      0\,, & \,\,\text{if}\,\,\,
      l_0^{(a)}< \f{c}{M}\,,\\
      1\,, & \,\,\text{if}\,\,\,
      l_0^{(a)}= \f{c}{M}\,.
    \end{aligned}
\right.
\end{equation}
Therefore, the contributions of states $\alpha$ containing particles
with non-maximal effective radius, i.e., with $l^{(a)}_0< \f{c}{M}$, 
are seen to be suppressed exponentially in $\chi$ when expressing
$C_n$ as a function of $z$, with factors of the form
$e^{-\delta\chi}z^\beta$ with $\delta$ and $\beta$ positive real
quantities, related to the masses and spin configuration of $\alpha$. 

In principle, it is possible that there are several particles for
which the ratio $l^{(a)}_0=\f{s^{(a)}-1}{m^{(a)}}$ is equal to the
maximum, Eq.~\eqref{eq:change_var_2}. For sure, if it is so for a
particle, so it is for its antiparticle. For simplicity, we will
assume that the maximum in Eq.~\eqref{eq:change_var_2} is essentially
unique, i.e., that there is a single particle-antiparticle pair that
saturates it (of course, particle and antiparticle may coincide);
the generalisation is straighforward, requiring only to take into
account the appropriate combinatorics. If $\tilde m$ and $\tilde s$
are respectively the mass and the spin of these particles, we can
conveniently choose $c=\tilde s-1$ and $M=\tilde m$.  
There are two possibilities.  

\paragraph{1.} Suppose that the relevant particle is a boson
coinciding with its antiparticle, and therefore having vanishing
discrete charges (baryon number, electric charge, ``strangeness'',
``charm'', ``bottomness'' and ``topness''). In this case, in the limit
of large $\chi$, the only terms that survive in the sum over $\alpha$
are the states $\alpha_n$ containing only $n$ such bosons. 
Since for these states ${\cal P}_{\alpha_n}=1$,
Eq.~\eqref{eq:not_G_3_an_con_large_chi_large_b_3} simplifies to 
\begin{equation}
  \label{eq:not_G_3_an_con_large_chi_large_b_4}
  \begin{aligned}
C_n(-i\chi;b;\nu_1,\nu_2) &   \mathop\sim_{\chi\to\infty} 
\left(\f{iw}{\sqrt{2\pi}}\right)^n
{\cal F}_{\alpha_n}(\{0\}_{\alpha_n};\nu_1)
    \overline{\cal F}_{\alpha_n}(\{0\}_{\alpha_n};\nu_2)
\\ &~\,\equiv \left(\f{iw}{\sqrt{2\pi}}\right)^nC_n^0(\nu_1,\nu_2)
\,,
  \end{aligned}
\end{equation}
where $w=w(\chi,z)$ is defined as
\begin{equation}
  \label{eq:def_of_w}
  w(\chi,z) \equiv z
\left[\log\left(\f{e^{\left(\tilde s
          -1\right)\chi}}{z}\right)\right]^{-\f{1}{2}}\,. 
\end{equation}

\paragraph{2.} Suppose that the relevant particle is a
fermion, not coinciding with its antiparticle, or a boson with
nonvanishing discrete charges.\footnote{The case of a
  fermion coinciding with its antiparticle, and the case of a boson
  not coinciding with its antiparticle but having vanishing
  discrete charges, are not relevant to QCD.} In this case, as the
selection rules on the discrete charges imply that only states with
vanishing baryon number, electric charge, etc., contribute to the sum
over $\alpha$, the only states that survive at large $\chi$ are
those containing only pairs of the relevant particle and
antiparticle. The total particle number must therefore be even,
$n=2k$, and the combinatorial factors of the relevant states
$\alpha_{2k}$ are equal to ${\cal
  P}_{\alpha_{2k}}=\f{(2k)!}{(k!)^2}$. Therefore, 
Eq.~\eqref{eq:not_G_3_an_con_large_chi_large_b_3} simplifies to   
\begin{equation}
  \label{eq:not_G_3_an_con_large_chi_large_b_6}
  \begin{aligned}
C_{2k}(-i\chi;b;\nu_1,\nu_2) & \mathop\sim_{\chi\to\infty} 
\left(\f{iw}{\sqrt{2\pi}}\right)^{2k}
 \f{(2k)!}{(k!)^2}
{\cal F}_{\alpha_{2k}}(\{0\}_{\alpha_{2k}};\nu_1)
    \overline{\cal F}_{\alpha_{2k}}(\{0\}_{\alpha_{2k}};\nu_2)
\\ &~\,\equiv
\left(\f{iw}{\sqrt{2\pi}}\right)^{2k}\f{(2k)!}{(k!)^2}\,C_{2k}^0(\nu_1,\nu_2) 
\,,
  \end{aligned}
\end{equation}
while the leading contribution to $C_{2k+1}$ must contain a boson 
of the type discussed above in point 1, with a nonmaximal
ratio $l_0^{(a)}$, and therefore is exponentially suppressed in
$\chi$ at fixed $w$. 

\paragraph{}
From the expressions above, it is immediate to see that $C_n$ depends
on $\chi$ and $b$ only through the factor $w^n$, independently of what
scenario is actually realised,\footnote{In case 2, this is true only
  for $n=2k$, while $C_{2k+1}$ is exponentially suppressed in $\chi$.}
up to subleading terms which are suppressed by at least one power of
$\chi$. In conclusion, we find that 
\begin{equation}
  \label{eq:cmb}
  \begin{aligned}
&    {\cal C}_M(\chi;\vec b_\perp;\nu_1,\nu_2)
    = \tilde{\cal C}_E(\theta \to -i\chi;
    b;\nu_1,\nu_2)  \\ &\phantom{uuu} \mathop\sim_{\chi\to\infty}
    ~~ g(w;\nu_1,\nu_2) - 1 \equiv \left\{
\begin{aligned}
&    \sum_{n=1}^{\infty}\f{1}{n!}\left(\f{iw}{\sqrt{2\pi}}\right)^n
 C_n^0(\nu_1,\nu_2)  &&& &\text{(case 1)}\,,\\ 
&    \sum_{k=1}^{\infty}\f{1}{(k!)^2}\left(\f{iw}{\sqrt{2\pi}}\right)^{2k}
 C_{2k}^0(\nu_1,\nu_2) 
&&& &\text{(case 2)}\,.    
    \end{aligned}
\right.
  \end{aligned}
\end{equation}
Here we have implicitly assumed that
${\cal F}_\alpha$ and/or $\overline{\cal F}_\alpha$ are nonzero at
$\{p_3\}_\alpha=\{0\}_\alpha$: the modifications to Eq.~\eqref{eq:cmb}
required when they vanish are discussed in Appendix \ref{app:1_1}.  
The bottom line is that at high energy, the dependence of the
correlator on $\chi$ and $b$ in the ``tail'' region $b>b_0$ [see
Eq.~\eqref{eq:condition}] is entirely encoded in the function
$w(\chi,z(\chi,b))$ defined above in Eqs.~\eqref{eq:change_var_0} and
\eqref{eq:def_of_w}. Going back to our discussion of effective radii,
Eqs.~\eqref{eq:not_G_3_an_con_large_chi_large_b_4},
\eqref{eq:not_G_3_an_con_large_chi_large_b_6} and \eqref{eq:cmb}
simply state that the large-$\chi$ behaviour is determined by the
particle(s) with the largest effective radius. The consequences of
this fact will be explored in the next Section. As a final remark, we
anticipate that the important feature of the result Eq.~\eqref{eq:cmb} 
is that in the high-energy limit the amplitude depends only on a
specific combination of $\chi$ and $b$. As we will see, this allows to
disentangle the energy dependence of the scattering amplitude at large
$\chi$. 

The validity of Eq.~\eqref{eq:cmb} relies mainly on the possibility of
interchanging the order in which one performs the sum over
intermediate states and the analytic continuation to Minkowski space,
and proving that this is actually allowed is currently out of reach in
the general case. It is however possible to provide a partial
justification, based on the short-range nature of strong
interactions. The basic observation is that the Wilson-loop matrix
elements $W_\alpha$ and $\overline W_\alpha$ in the limit of infinite
loop length, Eq.~\eqref{eq:def_w_al} and \eqref{eq:inf_T}, can be
written in factorised form to a first approximation. In the LSZ
framework~\cite{LSZ1,LSZ2}, $W_\alpha$ and $\overline W_\alpha$ are
obtained from the vacuum expectation value of the $T$-ordered product
of the Wilson loop and of appropriate interpolating local fields,
corresponding to each particle appearing in $\alpha$, integrated over
the position of the fields. Due to the finite interaction range, in
most of the configurations the interpolating fields will be far away
from each other, and therefore their mutual (``particle-particle'')
interactions will be negligible. Furthermore, they will interact with
the Wilson loop only locally (see footnote \ref{foot:LSZ}), so that
each of them will ``see'' in practice a loop of infinite length and
nothing else. The conclusion is that to first order one has
\begin{equation}
  \label{eq:factorised}
  \begin{aligned}
    W_\alpha(\{ \vec p\}_\alpha,
    \{s_3\}_\alpha;\nu_1) &\simeq \prod_{a,\,n_a(\alpha)\ne 0}
    \prod_{i=1}^{n_a(\alpha)} \lim_{T\to\infty}
    \f{\la 0 |\hat{\W}_E[\tilde{\C}^{\,(T)}_0(\nu_1)]
      | \alpha,\vec p^{\,(a)i},s_3^{(a)i}~;~in\ra }{ \la 0 |
      \hat{\W}_E[\tilde{\C}^{\,(T)}_0(\nu_1)] | 0\ra} \\ & \equiv 
    \prod_{a,\,n_a(\alpha)\ne 0}\prod_{i=1}^{n_a(\alpha)} W_{a}(\vec
    p^{\,(a)i},s_3^{(a)i};\nu_1)\,,
  \end{aligned}
\end{equation}
where $W_a$ are one-particle matrix elements, 
and similarly for $\overline W_\alpha$. From Eqs.~\eqref{eq:expansion}
and \eqref{eq:not_G_2}, and using the multinomial theorem, one finds 
\begin{equation}
  \label{eq:factorised2}
  \tilde C_E \simeq \exp\left\{ \sum_a
    \sum_{s_3=-s^{(a)}}^{s^{(a)}} e^{i\theta s_3} \int d\Omega_a \,
    e^{-b\varepsilon^{(a)}}
W_{a}({\cal R}_{\f{\theta}{2}}\vec p_a,s_3;\nu_1) \overline W_{a}({\cal
  R}_{-\f{\theta}{2}}\vec p_a,s_3;\nu_2)\right\} - 1\,,
\end{equation}
where $d\Omega_a=d^3p_a/[(2\pi)^32\varepsilon^{(a)}]$ is the
phase-space element for a particle of type $a$, and
$\varepsilon^{(a)}$ the corresponding energy.   
For simplicity, we have considered here only particles of the kind
discussed above in point 1, in order to avoid unnecessary
complications. For particles of the kind considered above in point 2,
the factorisation will be at the 2-particle level due to the selection
rules, and Eq.~\eqref{eq:factorised2} has to be modified to include
their contribution: this is discussed in Appendix \ref{app:1_2}.   
As the sum is now over the set of {\it asymptotic} particle species,
which is {\it finite} (at least in QCD), there are no more
complications due to problems of convergence of the sum, and one can 
safely perform the analytic continuation. Furthermore, one can
explicitly verify that the resummation can be done also if one
performs the analytic continuation first, and that this leads to the
same result. This proves that the term-by-term analytic continuation
is justified when particle-particle interactions can be 
neglected. Including the corrections due to particle-particle 
interactions will modify the expression above, but we think that it is
reasonable to assume that it will not spoil the possibility of
interchanging summation and analytic continuation. 

It is worth noting that even if the term-by-term analytic continuation
can be performed only for a limited range of $\chi$ at any fixed $b$,
which we expect to include higher and higher energies as we increase
the impact parameter,
Eq.~\eqref{eq:not_G_3_an_con_large_chi_large_b_3}  
shows that in this case we could nevertheless take the large-$\chi$
limit at fixed $w$, which amounts to take at the same time the
large-$\chi$ and the large-$b$ limit. This means that in this case
Eq.~\eqref{eq:cmb} would define the coefficients of a convergent power
series at least in some limited range of $w$. In this case, even
though the power series representation would be valid only within its
finite radius of convergence, it is possible that the analytic
function obtained by resumming the series could be analytically
continued (in $w$) outside the radius of convergence.  

More precisely, one can formulate the following condition. Let us
assume that the double series defined by Eqs.~\eqref{eq:expansion} and 
\eqref{eq:not_G_3}, re-expressed in terms of the {\it complex}
variable\footnote{Here we are assuming $\tilde s >1$.}
\begin{equation}
  \label{eq:w_comp_var}
  w = \f{e^{i\theta(\tilde s-1)}e^{-b\tilde m}}{\sqrt{b\tilde m}}
\end{equation}
and of $\theta$, is such that a term-by-term analytic continuation of
$\theta\to -i\chi$ at {\it fixed} $w$ can be performed, for $w\in
{\cal D}_0$ with ${\cal D}_0$ some complex domain. Possibly, the
analytic continuation has to be understood as $\theta\to
\epsilon-i\chi$, followed by the limit $\epsilon\to 0$. Let us assume
furthermore that in a subdomain ${\cal D}_1\subseteq {\cal D}_0$, that
we assume to contain part of the positive real axis in the complex-$w$
plane, at least as a boundary, it is possible to take
$\chi\to\infty$ at fixed $w$. Taking $w\in {\cal D}_1$, performing the
analytic continuation in $\theta$, and setting
\begin{equation}
  \label{eq:c_summary}
{\cal C}_M(\chi;\vec b_\perp;\nu_1,\nu_2)=  \tilde{\cal
  C}_E(-i\chi;b;\nu_1,\nu_2) = 
  F(w(\chi,z(b,\chi));\chi;\nu_1,\nu_2)\,, 
\end{equation}
according to the discussion above one has that $F$ can be written as
follows, 
\begin{equation}
  \label{eq:c_summary_2}
  F(w;\chi;\nu_1,\nu_2) =
g\left(w;\nu_1,\nu_2\right) +
  g_1\left(w;\chi;\nu_1,\nu_2\right)\,, 
\end{equation}
with $g_1\to 0$ as $\chi\to\infty$ at fixed $w$. Moreover, if $F$ and 
$g$ can be analytically extended beyond ${\cal D}_1$ including a
larger part (possibly all) of the real axis, then so can be $g_1$,
which will vanish as $\chi\to\infty$ in the whole extended domain.

\nsection{Elastic scattering amplitude and total cross section}
\label{sec:3}

We are now in a position to discuss the high-energy behaviour of the
meson-meson elastic scattering amplitude and of the corresponding
total cross section, so completing the argument outlined in Section
\ref{sec:outline}. For this purpose, it is useful to analyse first the
consequences of unitarity on the relevant Wilson loop correlators. 

It is clear from Eqs.~\eqref{eq:scatt-hadron} and
\eqref{eq:scatt-loop}, and from the definition of the scattering
amplitude in the impact-parameter representation, 
\begin{equation}
  \label{eq:imp_par_def}
  {\cal M}_{(hh)}(s,t) = -2 i s \int d^2\vec z_\perp\,  e^{i\vec q_\perp
    \cdot \vec z_\perp} a_{(hh)}(s,\vec z_\perp)\,,
\end{equation}
that $a_{(hh)}$ coincides with the Minkowskian Wilson-loop correlator
averaged over the dipole variables, i.e.,
\begin{equation}
  \label{eq:imp_par_1}
  a_{(hh)}(s,\vec z_\perp) =  \lla 
  {\cal C}_M(\chi;
  \vec z_\perp;\nu_1,\nu_2) \rra\,.
\end{equation}
It is well known~\cite{unitarity1,unitarity2,unitarity3} that the
impact-parameter amplitude satisfies the unitarity constraint $|
a_{(hh)}(s,\vec z_\perp) + 1|\le 1$, $\forall~\vec
z_\perp$. Therefore, if the description of the scattering process in
terms of dipoles, that we are using in this work, is to lead to
physically meaningful results, then the normalised Wilson-loop
correlator in Minkowski space has to satisfy the following 
unitarity constraint (at least in the large-$\chi$ limit),
\begin{equation}
  \label{eq:imp_par_2}
  | \lla {\cal C}_M(\chi;
  \vec z_\perp;\nu_1,\nu_2) \rra +1 | \le 1\,, \qquad \forall~\vec z_\perp\,.
\end{equation}
If the dipole picture is correctly describing {\it soft} high-energy
processes, and since the constraint Eq.~\eqref{eq:imp_par_2} has to be
satisfied for all the species of colliding mesons, i.e., for all the
physical choices of the wave functions $\psi_{1,2}$ in
Eq.~\eqref{eq:scatt-hadron}, we expect a stronger unitarity constraint
to be satisfied, namely 
\begin{equation}
  \label{eq:imp_par_2bis}
  | {\cal C}_M(\chi;\vec z_\perp;\nu_1,\nu_2)+1 | \le 1\,, \qquad
  \forall~\vec z_\perp,\nu_1,\nu_2\,. 
\end{equation}
In particular, choosing $\vec z_\perp=\vec b_\perp=(b,0)$ parallel to
the 2-axis, we have
\begin{equation}
  \label{eq:imp_par_3}
 | {\cal C}_M(\chi;\vec b_\perp;\nu_1,\nu_2) +1 | \le 1\,, \qquad
 \forall~b,\nu_1,\nu_2\,. 
\end{equation}
For $b>b_0$, where the analysis of the previous Section applies,
we have that ${\cal C}_M(\chi;\vec b_\perp)+1= g(w(\chi,z(\chi,b)))+
g_1(w(\chi,z(\chi,b));\chi)$, with $g_1\to 0$ as $\chi\to\infty$ with
$w$ fixed [see Eq.~\eqref{eq:cmb}]. Here we have dropped all the
irrelevant dependencies. The function $g$ is just the high-energy,
large-$b$ approximation of the normalised Wilson-loop correlator,
expressed as a function of $w$. As already remarked, one can keep $w$
fixed to any non-negative real value as $\chi\to\infty$ if one also
properly takes $b\to\infty$. We have therefore, according to
Eq.~\eqref{eq:imp_par_3}, 
\begin{equation}
  \label{eq:unitarity_g}
  |g(w;\nu_1,\nu_2)| = \lim_{\substack{\chi,b\to \infty \\ w
      ~{\rm fixed}}}
| {\cal C}_M(\chi; \vec b_\perp;\nu_1,\nu_2)+1 | \le 1 \qquad
\forall~ w\ge 0,\nu_1,\nu_2\,,
\end{equation}
i.e., $g$ is a bounded function.

\subsection{Asymptotic behaviour of the total cross section}
\label{subsec:asysigtot}

Recalling Eq.~\eqref{eq:rotations2_bis}, and exploiting the optical
theorem, we obtain for the total cross section  
\begin{equation}
  \label{eq:tot_cross}
  \begin{aligned}
    \sigma_{\rm tot}^{(hh)}(s) &\mathop{\sim}_{s \to \infty}  \frac{1}{s}
    \Im {\cal M}_{(hh)} (s, t=0)  =
    -4\pi \Re \,\lla \int_0^\infty db\, b\,
    {\cal C}_M(\chi ; \vec b_\perp;\nu_1,\nu_2) \rra_\varphi \,,
  \end{aligned}
\end{equation}
where $\vec b_\perp=(b,0)$ is parallel to the 2-axis, $\chi \simeq
\log(s/m^2)$ at high energy, and $\lla\ldots\rra_\varphi$ has been
defined in Eq.~\eqref{eq:rotations2_bis}. The integral in
Eq.~\eqref{eq:tot_cross} is conveniently split into two parts, 
\begin{equation}
  \label{eq:stot_split}
  \begin{aligned}
    \int_0^\infty db b\, {\cal C}_M(\chi ; \vec b_\perp;\nu_1,\nu_2)
    =&~ \int_0^{b_0(\nu_1,\nu_2)} db b\, {\cal C}_M(\chi ; \vec b_\perp
    ;\nu_1,\nu_2) \\ & +
    \int_{b_0(\nu_1,\nu_2)}^\infty db b\, {\cal C}_M(\chi ; \vec
    b_\perp;\nu_1,\nu_2)\,,  
  \end{aligned}
\end{equation}
where $b_0$ has been defined in Eq.~\eqref{eq:condition}. 
We expect from unitarity [see Eq.~\eqref{eq:imp_par_3}] that the first
term is bounded by a $\chi$-independent function, and so we will focus
on the second term. Here we can use the approximate expression for
${\cal C}_M$ obtained in the previous Section, Eq.~\eqref{eq:cmb}.   
Changing variables to $z$, as defined in the previous Section, i.e.,
\begin{equation}
  \label{eq:change_var}
  z = e^{(\tilde s - 1)\chi}e^{-\tilde m b}\,, \quad \tilde m b =
  \log\f{e^{(\tilde s - 1)\chi}}{z}\,, \quad \tilde m\, db = -\f{dz}{z}\,,
\end{equation}
and setting
$e^{-b_0\tilde m}=\Lambda$, with $\Lambda=\Lambda(\nu_1,\nu_2)$, we can write
\begin{equation}
  \label{eq:change_var_4}
  \begin{aligned}
 J(\chi;\nu_1,\nu_2)   &\equiv -\int_{b_0(\nu_1,\nu_2)}^\infty db b\,
 {\cal C}_M(\chi ; \vec b_\perp;\nu_1,\nu_2)\\
&  \mathop\sim_{\chi\to\infty}  
    \f{1}{\tilde m^2}\int_0^{e^{\left(\tilde s-1\right)\chi}\Lambda}
    \f{dz}{z} \, 
    \log\left(\f{e^{\left(\tilde s-1\right)\chi}}{z}\right)
   \left[1 - g(w(\chi,z);\nu_1,\nu_2)
    \right]  \,,
  \end{aligned}
\end{equation}
where $w(\chi,z)$ has been defined in Eq.~\eqref{eq:def_of_w}, and $g$
is defined in Eq.~\eqref{eq:cmb}. One could in principle use $w$
itself as integration variable, but for our purposes it is more
convenient to follow a different strategy. Let $z=\xi(\chi) z'$, and
let us require that $\xi$ is such that 
\begin{equation}
  \label{eq:change_var_2_bis}
w(\chi,z) = w(\chi,\xi(\chi) z') 
= \f{\xi(\chi)}{\sqrt{\log\left(\f{e^{\left(\tilde s
            -1\right)\chi}}{\xi(\chi)}\right)}} 
\f{  z'}{\sqrt{1+
    \f{\log\left(\f{1}{z'}\right)}{\log\left(\f{e^{\left(\tilde s
              -1\right)\chi}}{\xi(\chi)    
        }\right)} }}  \equiv \f{
  z'}{\sqrt{1+
    \f{\log\left(\f{1}{z'}\right)}{\log\left(\f{e^{\left(\tilde s
              -1\right)\chi}}{\xi(\chi)    
        }\right)} }}\,.
\end{equation}
The reason to do this is that now Eq.~\eqref{eq:change_var_4} depends
on $\chi$ only through the variable $\eta$, 
\begin{equation}
  \label{eq:ene_dep_var}
  e^\eta\equiv\f{e^{\left(\tilde s-1\right)\chi}}{\xi(\chi)}\,, \qquad
  \xi^2 +\log\xi = \left(\tilde s-1\right)\chi\,, \quad \eta =
  \left(\tilde s-1\right)\chi - \log\xi = \xi^2\,.
\end{equation}
The equation for $\xi$ in Eq.~\eqref{eq:ene_dep_var} can be solved,
yielding 
\begin{equation}
  \label{eq:asy_dep8_quat}
\eta=  \xi^2 = \f{1}{2} W(2e^{2\left(\tilde s-1\right)\chi})\,,
\end{equation}
where $W(z)$ is the Lambert $W$ function~\cite{LambertW}, defined by
the equation
\begin{equation}
  \label{eq:lambert_f}
  z = W(z)e^{W(z)}\,.
\end{equation}
For large positive $z$, $W(z) = \log z -\log\log z + \f{\log\log
  z}{\log z}+\ldots$, and so we obtain at large
energy
\begin{equation}
  \label{eq:asy_dep8_sex}
  \begin{aligned}
    \eta &= 
(\tilde s -1)\chi - \f{1}{2}\log[(\tilde s -1)\chi] + \f{\log
  [(\tilde s -1)\chi]}{4(\tilde s -1)\chi} + \ldots \,.
  \end{aligned}
\end{equation}
Dropping now the prime, rescaling $z\to \Lambda z$, and setting for
notational convenience
\begin{equation}
  \label{eq:other_defs}
\tilde g(z;\nu_1,\nu_2)= g(z\Lambda(\nu_1,\nu_2);\nu_1,\nu_2)\,,
\end{equation}
we can recast Eq.~\eqref{eq:change_var_4} in the compact form
\begin{equation}
  \label{eq:change_var_4_bis}
  \begin{aligned}
& J(\chi;\nu_1,\nu_2)   \mathop\sim_{\chi\to\infty}  
\f{1}{\tilde m^2}\int_0^{e^\eta}
    \f{dz}{z}
    \,\log\left(\f{e^\eta}{\Lambda z}\right)
    \Bigg(1 - \tilde g(z;\nu_1,\nu_2)
    \Bigg) \,,
  \end{aligned}
\end{equation}
where we have neglected terms of order ${\cal O}(\eta^{-1})$
appearing in Eq.~\eqref{eq:change_var_4} [see
Eq.~\eqref{eq:change_var_2_bis}]. Clearly, the constraint 
Eq.~\eqref{eq:unitarity_g} holds also for $\tilde g$.

In order to determine the high-energy behaviour of $J$, it is
convenient to split it into three parts, $J=J_1-J_2+J_3$, with
\begin{equation}
  \label{eq:J_split}
  \begin{aligned}
    J_1(\chi;\nu_1,\nu_2) &= \f{1}{\tilde m^2}\int_1^{e^\eta}
    \f{dz}{z}
    \,\log\left(\f{e^\eta}{\Lambda z}\right)\,,\\
    J_2(\chi;\nu_1,\nu_2) &= \f{1}{\tilde m^2}\int_1^{e^\eta}
    \f{dz}{z}
    \,\log\left(\f{e^\eta}{\Lambda z}\right)
    \tilde g(z;\nu_1,\nu_2)      \,,\\
      J_3(\chi;\nu_1,\nu_2) &= \f{1}{\tilde m^2}\int_0^{1}
    \f{dz}{z}
    \,\log\left(\f{e^\eta}{\Lambda z}\right)
    \Big(1 - \tilde g(z;\nu_1,\nu_2)    \Big)\,.
  \end{aligned}
\end{equation}
The first integral can be easily computed, and gives
\begin{equation}
  \label{eq:J_split_2}
  J_1(\chi;\nu_1,\nu_2) = \f{1}{\tilde m^2}\left[\f{1}{2}\eta^2 +
    \eta\log\f{1}{\Lambda}\right]\,.
\end{equation}
Moreover, the dependence of $J_3$ on $\eta$ is easily exposed,
\begin{equation}
  \label{eq:J_split_3}
  J_3(\chi;\nu_1,\nu_2) = \f{1}{\tilde m^2}\left[\eta\, c_1(\nu_1,\nu_2) 
    + c_2(\nu_1,\nu_2)\right]\,, 
\end{equation}
where $c_{1,2}$ are functions of the dipole variables only, so that
$J_3$ is subleading in $\eta$. Finally, using the unitarity constraint 
Eq.~\eqref{eq:imp_par_3}, or 
Eq.~\eqref{eq:unitarity_g}, we can bound
$J_2$, 
\begin{equation}
  \label{eq:J_split_4}
  |J_2(\chi;\nu_1,\nu_2)| \le J_1(\chi;\nu_1,\nu_2)\,,
\end{equation}
which leads to the following bound on the total cross section at high
energy, 
\begin{equation}
  \label{eq:J_split_bound}
  \sigma_{\rm tot}^{(hh)}(s)
\mathop\simeq_{s\to\infty} 4\pi \Re \lla J(\chi;\nu_1,\nu_2) \rra_\varphi
 \mathop\lesssim_{s \to \infty} 
4\pi\f{\eta^2}{\tilde m^2} \simeq
  4\pi\f{(\tilde s -1)^2}{\tilde m^2}\left(\log \f{s}{m^2}\right)^2\,,
\end{equation}
where we have used Eq.~\eqref{eq:asy_dep8_sex} and the fact that $\lla
1 \rra_\varphi = 1$. Notice that the subleading terms neglected in the last
passage are of order ${\cal O}(\log s \cdot \log\log s)$, as can be
seen again from Eq.~\eqref{eq:asy_dep8_sex}.  This ``Froissart-like''
bound is a consequence of the analyticity and unitarity properties
assumed for the Wilson-loop correlator, and of the existence of a
maximal (and positive) ratio $\f{(\tilde s -1)}{\tilde m}$ in the
hadronic spectrum. 
The origin of the prefactor is therefore rather different than in the
original derivation of the FLM bound, where it is related to the
position of the lowest singularity in the $t$-channel.

\subsection{Universality of total cross sections}

From Eqs.~\eqref{eq:J_split_2} and \eqref{eq:J_split_4}, we can
write for the asymptotic energy dependence of the total cross section 
\begin{equation}
  \label{eq:sigma_general}
  \sigma_{\rm tot}^{(hh)}(s) \mathop\sim_{s \to \infty}
2\pi\f{(\tilde s -1)^2}{\tilde m^2} \left[1  -
  \lla \Re\Delta(\nu_1,\nu_2) \rra_\varphi \right]
\left(\log \f{s}{m^2}\right)^2 \,, 
\end{equation}
where we have set 
$J_2 \sim
\f{1}{2}\Delta(\nu_1,\nu_2)\eta^2$ at high energy, compatibly with
Eq.~\eqref{eq:J_split_4}. Since the quantity $\Delta$ can in principle
depend nontrivially on the dipole variables, in general the resulting
total cross section will be nonuniversal. In order to have
universality, either $\Delta$ is purely imaginary (or even zero), or
$\Re\Delta$ is independent of $\nu_{1,2}$, or for some reason the
average of $\Re\Delta$ over the dipole variables is universal, which
would still be rather natural if only the average over the
orientations of the dipoles and over the momentum fractions were
required.  

To investigate the issue of universality, we exploit
Eq.~\eqref{eq:unitarity_g} to write 
\begin{equation}
  \label{eq:param_tg}
  \tilde g(z;\nu_1,\nu_2) = e^{-\rho(z;\nu_1,\nu_2)+i\phi(z;\nu_1,\nu_2)}\,,
\end{equation}
with $\rho,\phi\in\mathbb{R}$ and $\rho\ge 0$. Let us now consider
several interesting cases. 

\paragraph{1.} The simplest possibility is that
$\rho(z\to\infty)\to\infty$, in which case $\tilde g(z\to\infty) \to
0$, and we see that in Eq.~\eqref{eq:J_split} we can push the upper
limit of integration in $J_2$ to infinity, obtaining a finite
quantity.\footnote{This requires $\tilde g$ to vanish at least as fast
  as $|\log z|^{-2-\epsilon}$.} This implies that 
\begin{equation}
  \label{eq:J2_univ}
  \begin{aligned}
    J_2(\chi;\nu_1,\nu_2) =&~ \f{\eta}{\tilde m^2}\left[\int_1^{\infty}
    \f{dz}{z}
    \,\tilde g(z;\nu_1,\nu_2)
+ o(1)\right]  \\ &+ 
\f{1}{\tilde m^2}\int_1^{\infty}
    \f{dz}{z}\log\left(\f{1}{\Lambda z}\right)\,\tilde g(z;\nu_1,\nu_2)
+ o(1) 
\,,
  \end{aligned}
\end{equation}
i.e., $J_2={\cal O}(\eta)$ and therefore $\Delta=0$, so that
universality is obtained. In physical terms, this case corresponds
to the Minkowskian Wilson-loop correlator vanishing at large $\chi$
for any fixed $b$ (at least for any fixed $b>b_0$), so that in
this limit the impact-parameter amplitude goes to $1$. In other words,
this would directly correspond to the usual assumption of particles
behaving in a scattering process as black disks with energy-dependent
radius. 

\paragraph{2.} Another possibility we want to discuss is that of an
amplitude that keeps oscillating as $z\to\infty$, i.e., $\phi(z) \sim
\phi_0 z^\lambda$, $\lambda>0$, as $z\to\infty$. Separating the leading
contribution from the rest, setting $\tilde
g(z)=e^{-\rho(z)}e^{i(\phi(z)-\phi_0 z^\lambda)}e^{i\phi_0 z^\lambda}
= \tilde A(z) e^{i\phi_0 z^\lambda}$, and changing variables to
$y=z^\lambda$ in the integral for $J_2$, Eq.~\eqref{eq:J_split}, we
obtain 
\begin{equation}
  \label{eq:g_fourier}
  J_2(\chi;\nu_1,\nu_2) = \f{1}{\lambda^2\tilde m^2}\int_1^{e^{\lambda\eta}}
    \f{dy}{y}
    \,\log\left(\f{e^{\lambda\eta}}{\Lambda^\lambda y}\right)
   e^{i\phi_0 y}\tilde A(y^{\f{1}{\lambda}})\,.
\end{equation}
Since the function
$f_\lambda(y)=\f{1}{y}\log\left(\f{e^{\lambda\eta}}{\Lambda^\lambda
    y}\right)\theta(y-1)$ is ${\cal L}^2(\mathbb{R})$ and $\tilde A$ 
is bounded, we can push the upper limit of integration to infinity,
obtaining the Fourier transform of $f_\lambda \tilde A$ at
$\phi_0$. As a consequence, the leading behaviour of $J_2$ is only 
${\cal O}(\eta)$; therefore $\Delta=0$, and the colliding particles
behave effectively as black disks. Interestingly, we recover the same
result obtained in the previous case starting from completely
different assumptions. The key point is the possibility of pushing
the upper limit of integration to infinity in $J_2$, so that
Eq.~\eqref{eq:J2_univ} above holds: any time that this can be done, no
matter under what conditions, we will get $\Delta=0$ and universality
of the total cross sections. 

\paragraph{3.} The last case we want to discuss is that in which both
$\phi$ and $\rho$ become independent of $z$ at large $z$ while
remaining finite, i.e., $\phi(z\to\infty)=\phi_\infty(\nu_1,\nu_2)$,
and $\rho(z\to\infty)\to\rho_\infty(\nu_1,\nu_2)$. In this case, we
can write 
\begin{equation}
  \label{eq:uni_j2}
  J_2(\chi;\nu_1,\nu_2) = \f{1}{\tilde m^2}\int_1^{e^\eta}
    \f{dz}{z}
    \,\log\left(\f{e^\eta}{\Lambda
        z}\right)\left[e^{-\rho_\infty(\nu_1,\nu_2)}e^{i\phi_\infty(\nu_1,\nu_2)} +
      r(z;\nu_1,\nu_2)\right]\,, 
\end{equation}
where the quantity $r(z;\nu_1,\nu_2)\equiv
e^{-\rho(z;\nu_1,\nu_2)}e^{i\phi(z;\nu_1,\nu_2)}-
e^{-\rho_\infty(\nu_1,\nu_2)}e^{i\phi_\infty(\nu_1,\nu_2)}$ vanishes 
as $z\to\infty$. As a consequence, in the corresponding contribution
to the integral one can again push the upper limit of integration to 
infinity, obtaining a finite quantity.\footnote{This requires $r$ to
  vanish at least as fast as $|\log z|^{-2-\epsilon}$.} In turn, this
implies that this contribution grows at most like $\eta$ at large
energy. A leading contribution to $J_2$, proportional to $\chi^2$, is
however possible, and reads 
\begin{equation}
  \label{eq:uni_j2_2}
  J_2(\chi;\nu_1,\nu_2) =  \f{e^{-\rho_\infty(\nu_1,\nu_2)}}{2\tilde
    m^2} e^{i\phi_\infty(\nu_1,\nu_2)} \eta^2+ {\cal O}(\eta)\,.
\end{equation}
The leading contribution to the forward scattering amplitude is
therefore 
\begin{equation}
  \label{eq:uni_j2_3}
    {\cal M}_{(hh)}(s,0) = i s \,
\f{ 2\pi(\tilde s -1)^2 }{\tilde m^2} \chi^2 \left(1-
  e^{-\kappa^{(hh)}} e^{i\varphi^{(hh)}} 
\right) \,,
\end{equation}
where we have set
\begin{equation}
  \label{eq:uni_j2_4}
 e^{-\kappa^{(hh)}} e^{i\varphi^{(hh)}}= \lla
 e^{-\rho_\infty(\nu_1,\nu_2)} e^{i\phi_\infty(\nu_1,\nu_2)}  \rra_\varphi\,, 
\end{equation}
with $\kappa^{(hh)},\varphi^{(hh)}\in\mathbb{R}$ and $\kappa^{(hh)}
\ge 0$. In general, both $\kappa^{(hh)}$ and $\varphi^{(hh)}$ will
depend on the kind of particles involved in the scattering
process. However, not any value of $\varphi^{(hh)}$ is allowed, due to 
the constraint on the phase of scattering amplitudes at high energy
coming from analyticity and crossing symmetry (see, e.g.,
Ref.~\cite{Eden}). Due to this constraint, the crossing-symmetric part 
of the amplitude must be purely imaginary at high energy. If, for
simplicity, the hadronic wave functions are chosen to be invariant
under rotations and under the exchange $f_i\to 1-f_i$, as mentioned in
Section \ref{sec:1}, the scattering amplitude is automatically
crossing-symmetric, and therefore one must have $\varphi^{(hh)} =
0,\pi$. Since this must hold for all physical choices of the hadronic 
wave functions, the simplest way to achieve this is to have
$\phi_\infty(\nu_1,\nu_2) \equiv 0$, or $\pi$, independently of
$\nu_i$. This would ``naturally'' lead to universality of total cross
sections only if at the same time
$\rho_\infty(\nu_1,\nu_2)=\bar\rho_\infty$, independently of $\nu_i$:
\begin{equation}
  \label{eq:sigma_general_case2}
  \sigma_{\rm tot}^{(hh)}(s) \mathop\sim_{s \to \infty}
2\pi\f{(\tilde s -1)^2}{\tilde m^2} \left[1  \mp
  e^{-\bar\rho_\infty} \right]
\left(\log \f{s}{m^2}\right)^2 \,, 
\end{equation}
where the upper (lower) sign corresponds to
$\phi_\infty(\nu_1,\nu_2) = 0$ $(\pi)$. According to the sign, one
would have in this case either grey-disk ($\phi_\infty=0$) or
anti-shadowing~\cite{TT1,TT2} ($\phi_\infty=\pi$) behaviour. In particular,
if $\bar\rho_\infty=0$ one would have respectively no $\log^2 s$
contribution, or saturation of the unitarity limit. Although we cannot 
exclude that this scenario is realised, we find it rather unpleasing,
since it requires several extra conditions to be met in order to
achieve universality, and contains one parameter (i.e., $1  \mp
e^{-\bar\rho_\infty}$) that remains undetermined at this stage. In
the remaining part of this Section we will discuss only the first
two cases, with $\Delta=0$.   

\paragraph{}
If Eq.~\eqref{eq:J2_univ} holds, we can set for convenience
\begin{equation}
  \label{eq:g_asym_c1}
  \int_1^{\infty} \f{dz}{z} \tilde g(z;\nu_1,\nu_2) = -C(\nu_1,\nu_2) +
  c_1(\nu_1,\nu_2) + \log\f{1}{\Lambda} 
    \,,
\end{equation}
and write for $J$ 
\begin{equation}
  \label{eq:J_full}
  4\pi J(\chi;\nu_1,\nu_2) \mathop\sim_{s \to \infty}
 \f{2\pi}{\tilde m^2}\,\eta^2 +
\f{4\pi}{\tilde m^2} \lla C(\nu_1,\nu_2) \rra_\varphi \,\eta\,,
\end{equation}
where $\eta$ has been defined in Eq.~\eqref{eq:asy_dep8_quat}.  
From this we find that the total cross section behaves asymptotically
as 
\begin{equation}
  \label{eq:sigma_universal}
  \sigma_{\rm tot}^{(hh)}(s) \mathop\sim_{s \to \infty}
2\pi \f{(\tilde s -1)^2}{\tilde m^2} \left(\log\f{s}{m^2}\right)^2 -
2\pi \f{(\tilde s -1)}{\tilde m^2} \log\f{s}{m^2}\cdot\log\log\f{s}{m^2}
+ Q^{(hh)}\log\f{s}{m^2}\,,
\end{equation}
where we have used the large-$s$ behaviour of $\eta$ 
[see Eq.~\eqref{eq:asy_dep8_sex}], we have introduced the
(process-dependent) coefficient $Q^{(hh)}$ of the $\log s$ term, 
and discarded further subleading terms. There are three important
consequences of this expression.  
\begin{enumerate}
\item The leading energy dependence is of the form  $\sigma_{\rm
    tot}^{(hh)}(s) \mathop\sim B\log^2 s$ for ${s \to \infty}$, with {\it
    universal} $B$, i.e., independent of the colliding mesons. One
  easily sees that extending the calculation to the case of mesons with
  different masses this term remains unaffected.
\item The universal coefficient $B=2\pi \f{(\tilde s -1)^2}{\tilde
    m^2}$ can be entirely determined from the hadronic spectrum.
\item The first subleading correction in energy is proportional to 
  $\log s \cdot \log\log s$, and it is also universal.  
\end{enumerate}
All of the above remains true if extra powers of $b$ have to be
taken into account in the large-$b$ behaviour, as mentioned after
Eq.~\eqref{eq:not_G_3_an_con_large_chi_large_b_2}. More precisely, the
coefficient of the subleading $\log s \cdot \log\log s$ term gets an
extra (universal) factor, while the leading term remains unaltered. A
detailed discussion is reported in Appendix \ref{app:1_1}, where we also
make contact with the parameterisations discussed in Ref.~\cite{GMM}.  

Let us now briefly discuss the energy dependence of further subleading
corrections. The $o(1)$ terms in the first square bracket in
Eq.~\eqref{eq:J2_univ} are expected to vanish quite fast, and if they 
vanish at least as fast as $\eta^{-1}$, then they simply give a
constant contribution to the total cross section, plus vanishing
terms. The terms that have been neglected in the integrand passing from
Eq.~\eqref{eq:change_var_4} to Eq.~\eqref{eq:change_var_4_bis} are of
order ${\cal O}(\eta^{-1})$, and so can provide at most a contribution
of order ${\cal O}(\eta)$ to the total cross section. However, a
direct calculation shows that in the cases 1--3 discussed above they
only contribute to ${\cal O}(1)$. The terms that have been discarded
from Eq.~\eqref{eq:cmb} on are bounded [this follows from
Eqs.~\eqref{eq:imp_par_3} and \eqref{eq:unitarity_g}] and suppressed
by a factor $\left[\log\left(\f{e^{\eta}}{z}\right)\right]^{-1}$, so
they can yield at most a contribution of order ${\cal O}(\eta)$ to the
total cross section, which has to be included in $Q^{(hh)}$ in
Eq.~\eqref{eq:sigma_universal}. In any case,
Eq.~\eqref{eq:sigma_universal} above gives the leading contribution to 
the total cross section, in the case $\Re\Delta=0$. 

\subsection{Elastic scattering amplitude at high energy}

To conclude this Section, we want to briefly discuss what happens to
the elastic scattering amplitude if the conditions leading to
$\Delta=0$ are met. As these conditions correspond in practice to
particles behaving as black disks at high energy, one expects that the
usual results obtained in the black-disk model hold. For simplicity,
we will work with rotation-invariant wave functions, as appropriate
for example for unpolarised scattering processes. In this case,
recalling Eq.~\eqref{eq:simple-scatt}, the elastic scattering
amplitude reads   
\begin{equation}
  \label{eq:simple-scatt_2}
  {\cal M}_{(hh)}(s,t) = -4\pi i s \,\lla \int_0^\infty db b\,
  J_0(b\sqrt{-t})\,{\cal C}_M(\chi \simeq \log(s/m^2);
  \vec{b}_\perp;\nu_1,\nu_2) \rra_0\,.
\end{equation}
If $\tilde g(z)$ defined in Eq.~\eqref{eq:other_defs} vanishes
sufficiently fast at large $z$ (any power law will do), or if it
oscillates wildly, one can repeat the argument carried out above for
the total cross section, taking now into account the Bessel function
$J_0(b\sqrt{-t})$ appearing in Eq.~\eqref{eq:simple-scatt_2}. A simple
calculation then shows that to leading order 
\begin{equation}
  \label{eq:simple-scatt_asy}
    {\cal M}_{(hh)}(s,t) \mathop \sim_{s\to\infty} 
    4\pi i \, s \left(\f{\eta}{\tilde m}\right)^2\left(\f{\tilde
        m}{q\eta}\right)J_1\left(\f{q\eta}{\tilde m} 
    \right) =
4\pi i \,
    s \left(\f{\eta}{\tilde m}\right)^2 \f{J_1(\varrho)}{\varrho}
    \,,
\end{equation}
where we have set $\varrho=\varrho(\chi,q)\equiv\f{q\eta}{\tilde
  m}$. This is actually a {\it black-disk} scattering amplitude
with radius $\f{\eta}{\tilde m}$. A few remarks are in order.
\begin{enumerate}
\item The amplitude is purely imaginary at high energy, as expected
  from analyticity and crossing symmetry, and from the fact that it is 
  ${\cal M}_{(hh)}(s,t) \propto s$ up to logarithms (see, e.g.,
  Ref.~\cite{Eden}).  
\item The ratio
  \begin{equation}
    \label{eq:AKM_ratio}
    \f{{\cal M}_{(hh)}(s,t)}{{\cal M}_{(hh)}(s,0)} \mathop
    \sim_{s\to\infty}  \f{2 J_1(\varrho)}{\varrho} 
  \end{equation}
  depends on $s$ and $t$ only through the combination $\varrho\sim
  \mu^{-1}\sqrt{-t}\log s$, with $\mu=\tilde m/(\tilde s -
  1)$. Furthermore, the function  
  \begin{equation}
    \label{eq:AKM_ratio2}
    f(\tau) = \lim_{s\to\infty}  \f{{\cal M}_{(hh)}(s,-\mu^2\tau
      \chi^{-2})}{{\cal M}_{(hh)}(s,0)} = \f{2
      J_1(\sqrt{\tau})}{\sqrt{\tau}} 
  \end{equation}
  is entire of order $1/2$, in agreement with a theorem by Auberson,
  Kinoshita and Martin~\cite{AKM}.
\item At high energy, the elastic scattering amplitude is a 
  {\it universal} function of the form
  \begin{equation}
    \label{eq:univ_ampl}
    {\cal M}_{(hh)}(s,t) \sim is
    \log^2\f{s}{m_1 m_2} M\left(\sqrt{-t}\log\f{s}{m_1 m_2}\right)\,,   
  \end{equation}
  where we have straightforwardly extended the result to the case of
  colliding particles of different mass. 
\end{enumerate}
The well-known results of the black-disk model therefore hold: in
particular, one can show that\footnote{This requires to use the
  expression Eq.~\eqref{eq:simple-scatt_asy} in the whole range 
  $t\in[-\infty,0]$, which can be justified assuming that the
  small-$t$ region gives the dominant contribution to the elastic
  cross section.} $\sigma^{(hh)}_{\rm el}= \sigma^{(hh)}_{\rm tot}/2$,
and that the $B$-slope at zero transferred momentum, 
\begin{equation}
  \label{eq:univ4}
  {\cal B}(s) \equiv \f{d}{dt}\left[\log\f{d\sigma^{(hh)}_{\rm
        el}}{dt}\right]_{t=0}\,, 
\end{equation}
satisfies the relation $8\pi{\cal B}(s)=\sigma^{(hh)}_{\rm  tot}$. 
Moreover, the differential elastic cross section vanishes for $t_0$
satisfying $|t_0|({\eta}/{\tilde m})^2 = x_0^2$, where $x_0\simeq
3.83$ is the first zero of the function $J_1(x)$. Identifying $t_0$
with the position $t_{\rm dip}$ of the dip seen in differential
elastic cross sections, one expects that at high energy $|t_{\rm
  dip}|\sigma^{(hh)}_{\rm tot}= 2\pi x_0^2$. Notice that experimental
data give $\sigma^{(hh)}_{\rm el}/\sigma^{(hh)}_{\rm tot} \simeq 0.26$
and $8\pi{\cal B}(s)/\sigma^{(hh)}_{\rm  tot}\simeq 1.97$ at
$\sqrt{s}=7\,{\rm TeV}$, and $\sigma^{(hh)}_{\rm
  el}/\sigma^{(hh)}_{\rm tot} \simeq 0.27$ at $\sqrt{s}=8\,{\rm
  TeV}$~\cite{TOTEM1,TOTEM2,TOTEM3,TOTEM4}. Furthermore, recent
analyses~\cite{BDdD,CN} show that $|t_{\rm dip}|\sigma^{(hh)}_{\rm
  tot}/(2\pi x_0^2)$ is well above 1 up to LHC energies. If the
black-disk picture is correct, this indicates that the asymptotic
region is still far away from the energies presently available at
colliders. 

Finally, one can uncover the nature of the Pomeron singularity in the
complex angular momentum plane at high energy, by means of the Mellin
transform, 
\begin{equation}
  \label{eq:mellin}
  \begin{aligned}
    \bar{\cal M}_{(hh)}(\omega,t) 
&= \int_{m^2}^\infty \f{ds}{s}\, \left(\f{s}{m^2}\right)^{-\omega}{\cal M}_{(hh)}(s,t)
    \,.
  \end{aligned} 
\end{equation}
From Eq.~\eqref{eq:simple-scatt_asy} one easily obtains
\begin{equation}
  \label{eq:mellin3}
    \bar{\cal M}_{(hh)}(\omega,t) 
=  \f{4\pi
  i\left(\f{m}{\mu}\right)^2}{\left[\left(\omega-1\right)^2
 - \f{t}{\mu^2}\right]^{\f{3}{2}} 
  }  \,,
\end{equation}
which already appeared in Ref.~\cite{GNL}. This shows that the Pomeron
singularity at positive $t$ is not a pole, but rather an algebraic
singularity, while at $t=0$ it is a triple pole. The position of the
singularity is 
\begin{equation}
  \label{eq:pomeron_sing}
  \omega_{\mathbb{P}}^\pm = 1 \pm \f{\sqrt{t}}{\mu}\,,
\end{equation}
i.e., the Pomeron trajectory is nonlinear.

\nsection{Conclusions}
\label{sec:c}

In this paper we have shown how to obtain the leading energy
dependence of hadronic total cross sections, in the framework of the
nonperturbative approach to {\it soft} high-energy scattering based on
Wilson-loop correlation
functions~\cite{DFK,Nachtmann97,BN,Dosch,LLCM1}, if certain nontrivial
analyticity assumptions are satisfied. The total cross sections turn
out to be of ``Froissart'' type, $\sigma_{\rm tot}^{(hh)}(s)
\mathop\sim B\log^2 s$ for ${s \to \infty}$. We have also discussed a
few scenarios in which the coefficient $B$ turns out to be universal,
i.e., independent of the hadrons involved in the scattering process.

Our results rely mainly on the possibility of expressing the
Wilson-loop correlator, at high energy and large impact parameter $b$,
as a function of the combination 
\begin{equation}
  \label{eq:w_concl}
 w(\chi,b)=e^{\chi(\tilde s -1)}\f{e^{-b\tilde
  m}}{\sqrt{b\tilde m}}\,, 
\end{equation}
where $\chi \simeq \log(s/m^2)$, and $\tilde s$ and $\tilde m$ are
respectively the spin and mass of the particle that maximises the
ratio $l_0^{(a)}={(s^{(a)}-1)}/{m^{(a)}}$ over the asymptotic
one-particle states $a$. 

The ``natural'' scenarios leading to universality of $B$ depend on the
large-$w$ behaviour of the Wilson-loop correlator (WLC). We have
discussed three possibilities: (1) WLC $\to 0$ as $w\to\infty$, (2) WLC 
oscillates as $w\to\infty$, and (3) WLC $\to {\rm const.}$ as
$w\to\infty$. In cases 1 and 2, $B$ turns out to be entirely
determined by the hadronic spectrum, and reads $B_{\rm
  th}^{(1,2)}=\f{2\pi}{\mu^2}$, where $\mu=\tilde m/(\tilde s -
1)$. In case 3, analyticity and crossing symmetry require that the
constant is real, and one finds that $B_{\rm th}^{(3)}=\kappa B_{\rm
  th}^{(1,2)}$, with $0\le\kappa\le 2$ due to unitarity; $\kappa$ is
however not determined at this stage. 

Although the precise form of $w(\chi,b)$ in Eq.~\eqref{eq:w_concl} 
depends on some technical assumptions on certain matrix elements of
the relevant Wilson loops, a more general form 
still holds if these assumptions are
relaxed, namely $\bar w(\chi,b)=e^{\chi
  c}e^{-bM}/(bM)^{(1+\lambda)/2}$. In this more general case, one
finds $B^{\rm (1,2) \,gen}_{\rm 
  th}=\f{2\pi c^2}{M^2}$, which is independent of $\lambda$; one
however loses the simple connection with the spectrum. We note in
passing that the possibility to express the Wilson-loop correlator as
a function of $\bar w(\chi,b)$ for some values of $c$, $M$ and
$\lambda$ can be made into a general assumption, independently of our
derivation: this would obviously lead to the same results discussed in
this paper. 

In our calculation, we found that the first subleading correction in
energy is proportional to $\log s \cdot \log\log s$.  
The approach to {\it soft} high-energy scattering based on Wilson-loop 
correlation functions is expected to correctly yield the leading
energy dependence of scattering amplitudes and total cross sections,
while it is not clear how trustworthy the subleading terms
are: indeed, to settle this question one should carefully estimate the
energy dependence of the subleading terms discarded in the derivation
of Eqs.~\eqref{eq:scatt-hadron} and \eqref{eq:scatt-loop} (see
Refs.~\cite{Nachtmann91,DFK,Nachtmann97,reggeon}). 
Nevertheless, a $\log s \cdot \log\log s$ term has never been
considered so far in fits to the experimental data, and we believe it
worth to include it in a systematic analysis of total cross
sections. Such an analysis is however beyond the scope of this paper:
we have only checked that in fits limited to the high energy region
only  ($\chi \gtrsim 5\div 7$) the resulting value of $B$ is not very
much affected by its presence.\footnote{The function used for our
  checks is of the form $\sigma_{\rm tot}^{(hh)} = B \eta^2 + C \eta +D$,
  approximating $\eta$ using its large-$\chi$ expansion,
  Eq.~\eqref{eq:asy_dep8_sex}.} Indeed, $B$ slightly increases, as
expected, when this extra term is included, but the change is smaller
than the errors. Our expectation is therefore that the value of $B$
would not change much also in a systematic analysis, so that we can
compare the theoretical prediction for $B$ with the value currently
reported by the Particle Data Group~\cite{pdg},  
\begin{equation}
  \label{eq:bexp}
B_{\rm exp} = \f{2\pi}{M^2}\,, \qquad M=3.04(3)~{\rm GeV}  \,,
\end{equation}
i.e., $B_{\rm exp} \simeq 0.680(13)~{\rm GeV}^{-2}$ [$0.2646(49)~{\rm
  mb}$].   

\begin{figure}[t]
  \centering
  \includegraphics[width=0.89\textwidth]{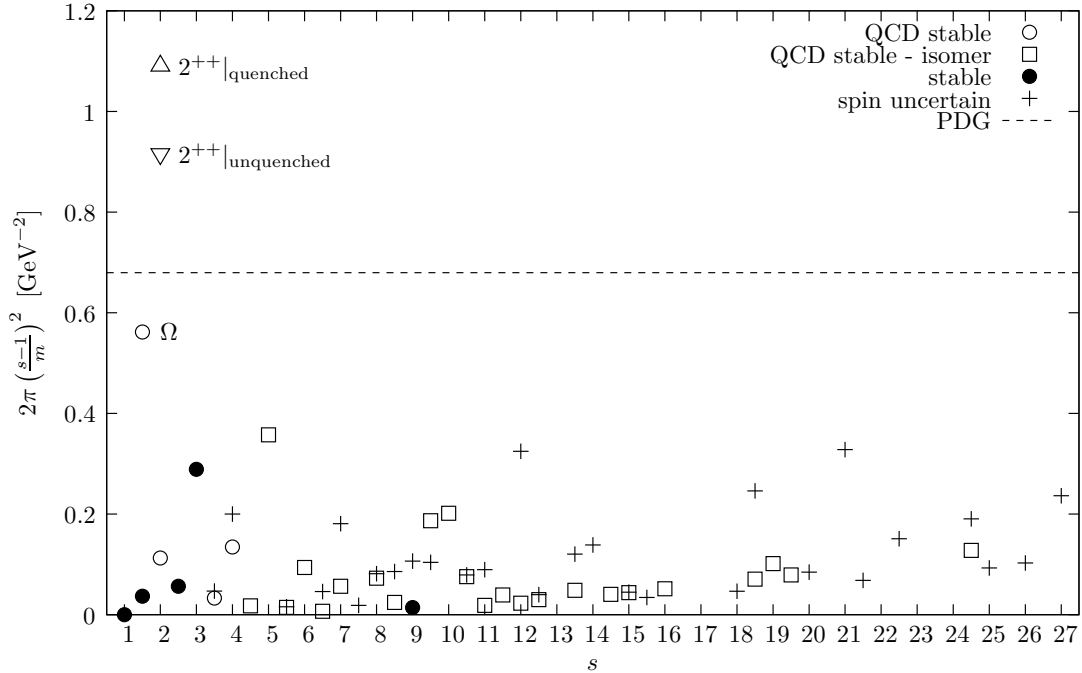}
  \caption{Value of the ratio $2\pi\left(\f{s-1}{m}\right)^2$ for
    QCD-stable one-particle states, plotted against the spin $s$, for
    $s\ge 1$. Only light mesons and baryons, and ground states and
    long-lived isomers of nuclei have been considered, and only the
    lightest states are shown for a given spin. Nuclear data are taken
    from Ref.~\cite{ABBW}. The maximal value of 
    $2\pi\left(\f{s-1}{m}\right)^2$ provides the coefficient
    $B_{\rm th}$ of the $\log^2 s$ term in $\sigma_{\rm tot}^{(hh)}$
    in the scenarios 1 and 2 discussed in the text. The value of
    $B_{\rm exp}$ reported by the Particle Data Group~\cite{pdg}, and
    the value of $B_{\rm th}$ obtained from the $2^{++}$
    {\it glueball} (data taken from Refs.~\cite{MP,GILMRRR}), are also
    shown for comparison.} 
  \label{fig:3}
\end{figure}

As we have said above, the theoretical expectation for $B$ in our
``favourite'' scenarios 1 and 2 is obtained from the spectrum of
stable particles with spin larger than 1, but one has to clarify what
``stable'' means in this context. As only strong interactions have
been considered in the derivation of the basic formula for hadronic
scattering amplitudes, Eqs.~\eqref{eq:scatt-hadron} and
\eqref{eq:scatt-loop}, a state has to be considered ``stable'' if it
is so when considering QCD in isolation. Among mesons and baryons with 
baryonic number 1, only the $\Omega^\pm$ is stable and with large
enough spin. The $\Omega^\pm$ baryon ($m_{\Omega^\pm}\simeq 1.67~{\rm
  GeV}$) has quantum numbers 
$J^P=\f{3}{2}^+$, electric charge $|Q|=1$ and strangeness $|S|=3$. The
other known QCD-stable states of high spin are nuclear states. A 
comprehensive study of nuclei is beyond the scope of this paper: here
we limit ourselves to the nuclear ground states and
long-lived isomers\footnote{A detailed study
  should include all the excited nuclear states that are QCD-stable.}
that are stable in Nature or decay through the electroweak
interactions. As can be seen in Fig.~\ref{fig:3} (data for nuclear
states are taken from Ref.~\cite{ABBW}), the $\Omega^\pm$ maximises
the relevant ratio, and yields $B_{\rm th}\simeq 0.56~{\rm GeV}^{-2}$ 
($0.22~{\rm mb}$), which is in fair agreement with experiments, being
about 20\% smaller than the value Eq.~\eqref{eq:bexp} reported by the 
Particle Data Group. In the comparison one should take into account
that the values of $B_{\rm exp}$ reported in
Refs.~\cite{Blogs1,Blogs2,Blogs3,Blogs4,Blogs5,Blogs6,Blogs7,Blogs8,pdg} 
show a variation of around 10\% due to different fitting procedures, 
which suggests a corresponding systematic error.

Strictly speaking, in this paper we have discussed meson-meson
scattering, starting from dipole-dipole scattering, while experimental
data for total cross sections are available mainly for baryon-baryon
and meson-baryon scattering. However, adopting a quark-diquark picture
for baryons (see Refs.~\cite{DFK,Nachtmann97,BN,Dosch,LLCM1,DR}), our
results extend immediately to these processes. Moreover, the arguments
of Section \ref{sec:2} can be easily generalised to more complicated
Wilson loops, aimed at describing the three-body structure of baryons,
or the inclusion of gluons and sea quarks in the description of
hadrons. Under the same assumptions made in this paper, one arrives at
the same behaviour of total cross sections obtained here, i.e., 
``Froissart-like'' total cross sections; also, the same considerations
can be made concerning universality. 
 
It is also worth noting, at this point, that the description of
hadrons in terms of dipoles is (probably) most naturally justified in
the {\it quenched} limit (or in the limit of a large number of colours
$N_c$), which would lead to consider the {\it quenched} (i.e.,
pure-gauge) theory as the relevant one. In this case, the relevant
spectrum would be the {\it glueball} spectrum, and, in particular, the
spectrum of stable {\it glueballs} with spin larger than 1, which,
according to Ref.~\cite{MP}, are those with $J^{PC} = 2^{++}$,
$2^{+-}$, $2^{-+}$, $2^{--}$, $3^{++}$, $3^{+-}$, $3^{--}$. Always
according to Ref.~\cite{MP} (but see also Refs.~\cite{Ky,Meyer} for
other more recent {\it quenched} determinations of the {\it glueball}
spectrum), among the {\it glueballs} with $J=2$, the
lightest (and thus relevant) one is the $2^{++}$, with
$M^{(Q)}_{2^{++}} \simeq 2.40~{\rm GeV}$, while the lightest {\it
  glueball} with $J=3$ is the $3^{+-}$, with $M^{(Q)}_{3^{+-}} \simeq
3.55~{\rm GeV}$.\footnote{\label{foot:glue} We would like to point
  out, at this point, that the identification of the spin of {\it
    glueball} states on the lattice is highly nontrivial and, in some
  cases (such as $2^{+-}$, $2^{--}$, $3^{++}$ and $3^{--}$), also
  controversial~\cite{MP}. Moreover, a more recent ({\it unquenched})
  determination of the {\it glueball} spectrum~\cite{GILMRRR} has {\it
    not} found evidence for the states $2^{+-}$, $3^{++}$ and
  $3^{--}$, leaving, however, the states $2^{++}$ and $3^{+-}$ as
  possible relevant states, with masses $M_{2^{++}} \simeq 2.62~{\rm
    GeV}$ and $M_{3^{+-}} \simeq 3.85~{\rm GeV}$.}  
They would lead to a value of the coefficient $B$ given by,
respectively, $B^{(Q)}_{2^{++}} \simeq 1.09~{\rm GeV}^{-2}$
($0.42~{\rm mb}$) and $B^{(Q)}_{3^{+-}} \simeq 1.99~{\rm GeV}^{-2}$
($0.77~{\rm mb}$). Therefore, according to our approach, we should
conclude that, in the {\it quenched} theory, 
$B^{(Q)}_{\rm th} = B^{(Q)}_{3^{+-}} \simeq 1.99~{\rm GeV}^{-2}$
($0.77~{\rm mb}$), which is (quite surprisingly) a factor 3 larger
than $B_{\rm exp}$. 

An interesting issue is the possible effectiveness of one-particle
selection rules to reduce the set of states over which $l_0^{(a)}$ has
to be maximised. It is rather easy to prove selection rules for spin 
($s_3$), parity ($\eta_P$) and charge conjugation ($\eta_C$) for the
one-particle contributions: they are nonzero only if $\eta_P = \eta_C
= e^{i\pi s_3}$. But even if a particle does not contribute at this
level, there is no reason for it not to contribute starting from the
two-particle level, so that it should be included in the set over
which $l_0^{(a)}$ has to be maximised. Things could change if the
(connected) contributions of many-particle states were suppressed  by
additional powers of $e^\chi$, but we have not found any argument
supporting this possibility. If, for some reason, states for which
one-particle contributions are nonzero should be considered to be
``dominant'', then (see also footnote \ref{foot:glue}) we would be left
with the {\it glueball} $2^{++}$, with $B^{(Q)}_{\rm th} =
B^{(Q)}_{2^{++}} \simeq 1.09~{\rm GeV}^{-2}$ ($0.42~{\rm mb}$), as it
has been already suggested in Ref.~\cite{Moriond}, commenting the
results of the best fits to the lattice data performed in
Ref.~\cite{GMM}.\footnote{We note in passing that a Pomeron of gluonic
  nature and effectively of spin 2 has been recently proposed in
  Ref.~\cite{EMN}.} This value is still larger than $B_{\rm
  exp}$, but ``only'' by a factor 1.6 [and an even better agreement
with $B_{\rm exp}$ would be obtained if we used the {\it unquenched}
value $M_{2^{++}} \simeq 2.62~{\rm GeV}$ found in Ref.~\cite{GILMRRR},
which leads to $B_{2^{++}} \simeq 0.91~{\rm GeV}^{-2}$ ($0.35~{\rm
  mb}$)]. Therefore, a cautious conclusion could be that, in the {\it
  quenched} theory, $B^{(Q)}_{\rm th}$ is {\it at least} $1.09~{\rm
  GeV}^{-2}$ ($0.42~{\rm mb}$). The comparison with the {\it
  unquenched} estimate $B_{\rm th} \simeq 0.56~{\rm GeV}^{-2}$
($0.22~{\rm mb}$), that we have found above,\footnote{Since the
  $\Omega^\pm$ baryon does not satisfy the selection rules on baryon
  number, electric charge and strangeness, the above-mentioned
  (hypothetical) suppression mechanism would make this value into an
  upper bound.} seems to suggest (quite surprisingly) large {\it
  unquenching} effects due to the {\it sea} quarks. Of course, also
the possibility that the relevant state (which maximises the ratio
$l^{(a)}_0 = \f{s^{(a)}-1}{m^{(a)}}$) has not yet been discovered,
cannot be excluded.  

\section*{Acknowledgements}

M.G. is supported by the Hungarian Academy of Sciences under
``Lend\"ulet'' grant No. LP2011-011.

\appendix

\nsection{Technical details}
\label{app:1}

\subsection{Vanishing Wilson-loop matrix elements and massless
  particles} 
\label{app:1_1}
In this Appendix we briefly discuss the consequences of 
vanishing matrix elements ${\cal F}_\alpha(\{0\}_\alpha;\nu_1)$ and/or
$\overline{\cal 
  F}_\alpha(\{0\}_\alpha;\nu_2)$ on the large-$\chi$ behaviour of the
relevant Wilson-loop correlator, and on the asymptotic behaviour of
total cross sections. Furthermore, we discuss the possible effects due
to the presence of massless particles in the spectrum. 

Let us assume that 
\begin{equation}
  \label{eq:extra_pow}
 {\cal
  F}_\alpha(\{p_3\}_\alpha;\nu_1)\overline{\cal
  F}_\alpha(\{p_3\}_\alpha;\nu_2) \simeq k_\alpha
\prod_{a,\,n_a(\alpha)\ne 0} \prod_{i=1}^{n_a(\alpha)}
\left(p_3^{(a)i}\right)^{\lambda_\alpha^{(a)}} \,,
\end{equation}
with $k_\alpha$ some function of $\nu_i$, and where some of the
$\lambda_\alpha^{(a)}$ can be zero. One immediately finds after the
change of variables Eq.~\eqref{eq:large_b1} that in the limit of large
$b$ the contribution $\delta C_{\alpha}$  of state $\alpha$ to $C_n$
is proportional to  
\begin{equation}
  \label{eq:extra_pow_2}
\delta C_{\alpha} \propto \delta_{{\cal N}_\alpha,n}
\prod_{a}\left(
\f{1}{(bm^{(a)})^{\f{1+\lambda_\alpha^{(a)}}{2}}} 
      e^{\chi[s^{(a)}-1]}e^{-b m^{(a)}}\right)^{n_a(\alpha)}\,.
\end{equation}
Here $\lambda_\alpha^{(a)}$ is the same for all particles of the same 
species due to symmetry reasons. As anticipated in Section
\ref{subsec:largeb}, the only effect of vanishing ${\cal F}_\alpha$, 
$\overline{\cal F}_\alpha$ on the large-$b$ behaviour of $\delta
C_{\alpha}$ is the appearence of extra inverse powers of $b$.  

The quantities ${\cal F}_\alpha(\{0\}_\alpha;\nu_1)$ and $\overline{\cal
  F}_\alpha(\{0\}_\alpha;\nu_2)$ are actually expected to vanish in
the presence of massless particles, since in this case the phase-space 
measure contains factors $dp_3/p_3$, and the integral would diverge
otherwise. Including massless particles requires only a minor
modification to the calculation of Section \ref{subsec:largeb}, 
and we obtain for $\delta C_{\alpha}$ in the limit of large $b$ 
\begin{equation}
  \label{eq:extra_pow_3}
\delta C_{\alpha} \propto \delta_{{\cal N}_\alpha,n} \prod_{a}\left(
\f{1}{(bm^{(a)})^{\f{1+\lambda_\alpha^{(a)}}{2}}} 
      e^{\chi[s^{(a)}-1]}e^{-b m^{(a)}}\right)^{n_a(\alpha)}
\prod_{a'}\left( \f{1}{b^{\lambda_\alpha^{(a')}}} 
      e^{\chi[s^{(a')}-1]}\right)^{n_{a'}(\alpha)}
\,,
\end{equation}
where the indices $a$ and $a'$ run over massive and massless
particles, respectively.

The modifications discussed above affect the asymptotic expression
Eq.~\eqref{eq:cmb} for the relevant Wilson-loop correlator.
If only massive particles are present, one has simply to introduce the
extra powers of $b$ discussed above, finding
\begin{equation}
  \label{eq:conn_large_chi_mod}
    C_n(-i\chi;b;\nu_1,\nu_2) 
\mathop\sim_{\chi\to\infty}  
    \f{(iz)^n}{\left[2\pi\log\left(\f{e^{\left(\tilde s
                -1\right)\chi}}{z}\right)^{1+\lambda}  
\right]^{\f{n}{2}}}\,\bar C_n^0(\nu_1,\nu_2)
\equiv \left(\f{iw_\lambda}{\sqrt{2\pi}}\right)^n\,\bar C_n^0(\nu_1,\nu_2)
\,,
\end{equation}
with $\lambda$ being the appropriate power corresponding to the
relevant particle(s).\footnote{For simplicity, we are assuming that
  $\lambda$ is $n$-independent. If $\lambda$ depends on $n$, only
  those terms with smallest $\lambda$ have to be kept to leading order
in $\chi$. Also, Eq.~\eqref{eq:conn_large_chi_mod} holds only for
$n=2k$ if we are in case 2 discussed in Section \ref{subsec:largechi},
while $C_{2k+1}$ is suppressed exponentially in $\chi$.} On the other
hand, in the presence of massless particles, one has 
to reconsider the procedure leading to Eq.~\eqref{eq:cmb}. If a state
contains massless particles, its contribution to $C_n$ takes the form
Eq.~\eqref{eq:extra_pow_3}. It is easy to see that if $(\tilde s 
-1)/\tilde m$ can be maximised over the massive particles (yielding
$\tilde s >1$), changing variables to $z$,
Eq.~\eqref{eq:change_var}, one gets extra factors from each massless 
particle of the form
\begin{equation}
  \label{eq:massless_contr}
 \left(\f{e^{\chi(s^{(0)}-1)}}{b^{\lambda^{(0)}}}\right)^{n^{(0)}}
= \left(\f{e^{\chi(s^{(0)}-1)}}{\left[\f{1}{\tilde
        m}\left(\chi(\tilde s-1) - \log
        z\right)\right]^{\lambda^{(0)}}}\right)^{n^{(0)}} 
\,.
\end{equation}
If $s^{(0)}=0$, Eq.~\eqref{eq:massless_contr} vanishes exponentially
in $\chi$ at high energy, meaning that scalar massless particles can
be safely neglected; this allows to safely consider the chiral limit
of QCD. If $s^{(0)}=1$, on the other hand,
Eq.~\eqref{eq:massless_contr} vanishes only as a power of $\chi$, so
that it can give important subleading contributions.\footnote{In
  principle these contributions could also be dominant, if
  $\lambda^{(0)}$ were smaller than the power corresponding to the
  relevant massive particles.} 
The situation is drastically different
if massless particles with $s^{(0)}\ge 2$ are present: in this case
the proper change of variable is rather
$z={e^{\chi(s^{(0)}-1)}}/{b^{\lambda^{(0)}}}$, with $s^{(0)}$ and 
$\lambda^{(0)}$ corresponding to the massless particle maximising the
ratio $(s-1)/\lambda$, which kills all the massive contributions, and
all the other massless contributions as well. We will not consider
this case any longer. 

The modifications discussed above have only mild consequences on the
asymptotic behaviour of the total cross section. To see this, one has
simply to repeat the calculation of Section \ref{subsec:asysigtot}, 
taking into account that now ${\cal C}_M(\chi
;\vec{b}_\perp;\nu_1,\nu_2) \sim \bar g(w_\lambda;\nu_1,\nu_2)-1$ for
$\chi\to\infty$, where 
\begin{equation}
  \label{eq:mod_w}
   w_\lambda(\chi,z) = z
\left[\log\left(\f{e^{\left(\tilde s
          -1\right)\chi}}{z}\right)\right]^{-\f{1+\lambda}{2}}\,,
\end{equation}
$z= e^{(\tilde s - 1)\chi}e^{-\tilde m b}$ is the same as in
Eq.~\eqref{eq:change_var}, and $\bar g$ is a bounded function, which
can be proved exploiting unitarity as in Eq.~\eqref{eq:unitarity_g}.  
Rescaling $z=\xi(\chi) z'$, with $\xi(\chi)$ defined by
\begin{equation}
  \label{eq:change_var_2_bis_mod}
w_\lambda(\chi,z) = w_\lambda(\chi,\xi(\chi) z') 
\equiv \f{
  z'}{\left[1+
    \f{\log\left(\f{1}{z'}\right)}{\log\left(\f{e^{\left(\tilde s
              -1\right)\chi}}{\xi(\chi)    
        }\right)} \right]^{\f{1+\lambda}{2}}}\,,
\end{equation}
setting again $\eta=\log\left(\f{e^{\left(\tilde s
        -1\right)\chi}}{\xi(\chi)    
  }\right)$, and solving the equation for $\xi$, one finds
\begin{equation}
  \label{eq:mod_eta}
\eta= \tilde \lambda
W\left(\f{1}{\tilde\lambda}e^{\f{1}{\tilde\lambda}\chi(\tilde 
  s-1)}\right)\,, 
\end{equation}
with $\tilde \lambda= (1+\lambda)/2$. One then proceeds as in Section
\ref{subsec:asysigtot}, and
if $\Delta=0$ [see Eq.~\eqref{eq:sigma_general}] one again obtains
Eq.~\eqref{eq:J_full}, with $\eta$ defined now in
Eq.~\eqref{eq:mod_eta}. The leading term in the expansion of $\eta$
at large $\chi$ is unchanged, while the first subleading correction is
modified,
\begin{equation}
  \label{eq:mod_X}
\eta = (\tilde s -1)\chi - \f{1}{2}\log[(\tilde s -1)\chi] \to (\tilde
s -1)\chi - \tilde 
  \lambda\log[(\tilde s -1)\chi]\,.
\end{equation}
As a consequence, also for nonzero $\lambda$ the total cross section
is of the form Eq.~\eqref{eq:sigma_universal}, with an extra factor of 
$\tilde\lambda$ in front of the coefficient of the subleading $\log s
\cdot \log\log s$ term, but with exactly the same leading term.

Finally, let us make contact with the parameterisations of lattice
data discussed in Ref.~\cite{GMM}. The functional forms considered
there for the Euclidean correlator, after analytic continuation to
Minkowski space and in the large-$\chi$ limit, reduce to the general
form 
\begin{equation}
  \label{eq:prev_pap}
  \begin{aligned}
    {\cal C}_M(\chi ;\vec{b}_\perp;\nu_1,\nu_2)
    &\mathop\simeq_{\chi\to\infty} \exp\{{\cal K}_M(\chi
    ;\vec{b}_\perp;\nu_1,\nu_2)\} -1\,, \\
    {\cal K}_M(\chi ;\vec{b}_\perp;\nu_1,\nu_2)
    &= i\beta(\nu_1,\nu_2) \chi^p e^{n\chi} e^{-\mu b}\,.
  \end{aligned}
\end{equation}
They are therefore functions of the variable $\tilde w = \chi^p \tilde
z$ only, with $\tilde z = e^{n\chi} e^{-\mu b}$, which, up to
subleading terms at large $\chi$, coincides with Eq.~\eqref{eq:mod_w}
upon identifying $n=\tilde s-1$, $\mu=\tilde m$ and
$p=-(1+\lambda)/2$. These parameterisations therefore lead to the same
high-energy behaviour of total cross sections found here, as
already discussed in Ref.~\cite{GMM}. Furthermore, if $\Im \beta >0$,
they satisfy our first criterion for universality, i.e., vanishing of
the Wilson-loop correlator at large $\chi$ and fixed $b$, while if
$\Im \beta =0$ the correlator oscillates wildly at large $\chi$ and
fixed $b$, thus satisfying our second criterion for universality. 

\subsection{Resummation for factorised matrix elements: general case}
\label{app:1_2}

We want now to extend the discussion of Section \ref{subsec:largechi}
concerning the possibility to interchange the order of summation and
analytic continuation to the case in which particles of type 2 (see
Section \ref{subsec:largechi}) are
present, when the Wilson-loop matrix elements are approximately in
factorised form. Including this kind of particles, the
factorised form of the matrix elements reads\footnote{Here we are
  assuming that a particle can be paired only to its antiparticle to
  evade the selection rules. Dropping this assumption would only make
  the combinatorics more complicated, without affecting the argument.}
\begin{equation}
  \label{eq:factorised_app}
  \begin{aligned}
    W_\alpha(&\{ \vec p\}_\alpha,
    \{s_3\}_\alpha;\nu_1) \simeq \\ & 
\prod_{a,\,n_a(\alpha)\ne 0}^{(1)}
    \prod_{i=1}^{n_a(\alpha)} \lim_{T\to\infty}
    \f{\la 0 |\hat{\W}_E[\tilde{\C}^{\,(T)}_0(\nu_1)]
      | \alpha,\vec p^{\,(a)i},s_3^{(a)i}~;~in\ra }{ \la 0 |
      \hat{\W}_E[\tilde{\C}^{\,(T)}_0(\nu_1)] | 0\ra} \\ & 
\times  \prod_{a,\,n_a(\alpha)\ne 0}^{(2)}
    \sum_{P_a} \prod_{i=1}^{n_a(\alpha)} 
    \lim_{T\to\infty}
    \f{\la 0 |\hat{\W}_E[\tilde{\C}^{\,(T)}_0(\nu_1)]
      | \alpha,\vec p^{\,(a)i},\vec
      p^{\,(\bar a)i_{P_a}},s_3^{(a)i},s_3^{(\bar a)i_{P_a}}~;~in\ra }{ \la 0 | 
      \hat{\W}_E[\tilde{\C}^{\,(T)}_0(\nu_1)] | 0\ra} \\ 
\equiv &
    \prod_{a,\,n_a(\alpha)\ne 0}^{(1)}\prod_{i=1}^{n_a(\alpha)} W_{a}(\vec
    p^{\,(a)i},s_3^{(a)i};\nu_1)\\ & \times
\prod_{a,\,n_a(\alpha)\ne 0}^{(2)}\sum_{P_a} \prod_{i=1}^{n_a(\alpha)}W_{a\bar{a}}(\vec
    p^{\,(a)i},\vec
      p^{\,(\bar a)i_{P_a}},s_3^{(a)i},s_3^{(\bar a)i_{P_a}};\nu_1)
\,,
  \end{aligned}
\end{equation}
where $W_a$ are one-particle matrix elements, and $W_{a\bar{a}}$ are
particle-antiparticle pair matrix elements; a similar result holds for
$\overline W_\alpha$. Here the superscript $(1)$ and $(2)$ indicate
that the products in Eq.~\eqref{eq:factorised_app} are restricted to
particles of type 1 and particle-antiparticle {\it pairs} of type 2,
respectively; $\bar a$ denotes the antiparticle of particle $a$; $P_a$
is a permutation of $1,\ldots,n_a(\alpha)$ [clearly
$n_a(\alpha)=n_{\bar a}(\alpha)$]. From
Eqs.~\eqref{eq:expansion} and \eqref{eq:not_G_2}, and using the
multinomial theorem, we get
\begin{equation}
  \label{eq:factorised2_app}
  \begin{aligned}
    \tilde C_E &\simeq \exp\left\{ \sum_a^{(1)}
    \sum_{s_3=-s^{(a)}}^{s^{(a)}} e^{i\theta s_3} \int d\Omega_a \,
    e^{-b\varepsilon^{(a)}}
W_{a}({\cal R}_{\f{\theta}{2}}\vec p_a,s_3;\nu_1) \overline W_{a}({\cal
  R}_{-\f{\theta}{2}}\vec p_a,s_3;\nu_2) \right. \\ &
\left. \phantom{\simeq \exp}~~
+ \f{1}{2}
\sum_a^{(2)}
    \sum_{s_3=-s^{(a)}}^{s^{(a)}} \sum_{\bar s_3=-s^{(a)}}^{s^{(a)}}
    e^{i\theta (s_3+\bar s_3)} \int d\Omega_ad\Omega_{\bar a} \, 
    e^{-b[\varepsilon^{(a)}+\varepsilon^{(\bar a)}]}
\right. \\ &
\left. \phantom{\simeq \exp\sum_a^{(2)}} \times
W_{a\bar a}({\cal R}_{\f{\theta}{2}}\vec p_a,{\cal
  R}_{\f{\theta}{2}}\vec p_{\bar a},s_3,\bar s_3;\nu_1) \overline
W_{a\bar a}({\cal 
  R}_{-\f{\theta}{2}}\vec p_a,{\cal 
  R}_{-\f{\theta}{2}}\vec p_{\bar a},s_3,\bar s_3;\nu_2)
\right\} - 1\,,
  \end{aligned}
\end{equation}
where $d\Omega_a=d^3p_a/[(2\pi)^32\varepsilon^{(a)}]$ is the
phase-space element for a particle of type 
$a$, and $\varepsilon^{(a)}$ the corresponding energy. The argument
then goes as in Section \ref{subsec:largechi}: the sums in the
exponent in Eq.~\eqref{eq:factorised2_app} are over finite sets, so
that there is no convergence problem, and one can verify explicitly
that analytic continuation and summation over the complete set of
states commute. 



\begin{thebibliography}{99}

\bibitem{TOTEM1}
  G.~Antchev {\it et al.} [TOTEM collaboration], Europhys. Lett. {\bf 96}
  (2011) 21002.

\bibitem{TOTEM2}
  G.~Antchev {\it et al.} [TOTEM collaboration],  Europhys.\ Lett.\
  {\bf 101} (2013) 21002.


\bibitem{TOTEM3}
  G.~Antchev {\it et al.} [TOTEM collaboration],  Europhys.\ Lett.\
  {\bf 101} (2013) 21004.

\bibitem{TOTEM4}
  G.~Antchev {\it et al.} [TOTEM collaboration],  Phys.\ Rev.\ Lett.\
  {\bf 111} (2013) 012001. 


\bibitem{KFK} 
  A.~K.~Kohara, E.~Ferreira and T.~Kodama,
  Eur.\ Phys.\ J.\ C {\bf 73} (2013) 2326
  [arXiv:1212.3652 [hep-ph]].

\bibitem{Dremin} 
  I.~M.~Dremin,
  JETP Lett.\  {\bf 97} (2013) 571
  [arXiv:1304.5345 [hep-ph]].

\bibitem{Dremin2} 
  I.~M.~Dremin and V.~A.~Nechitailo,
  Nucl.\ Phys.\ A {\bf 916} (2013) 241
  [arXiv:1306.5384 [hep-ph]].

\bibitem{GLM} 
  E.~Gotsman, E.~Levin and U.~Maor,
  Phys.\ Lett.\ B {\bf 716} (2012) 425
  [arXiv:1208.0898 [hep-ph]].



\bibitem{Blogs1}
  K.~Igi and M.~Ishida, Phys. Rev. D {\bf 66} (2002) 034023 
  [hep-ph/0202163].

\bibitem{Blogs2}
  J.~R.~Cudell {\it et al.} (COMPETE collaboration), Phys. Rev. D {\bf
    65} (2002) 074024 [hep-ph/0107219].

\bibitem{Blogs3}
  K.~Igi and M.~Ishida, Phys. Lett. B {\bf 622} (2005) 286 
  [hep-ph/0505058].

\bibitem{Blogs4}
  M.~M.~Block and F.~Halzen, Phys. Rev. D {\bf 72} (2005) 036006
  [Erratum-ibid.\ D {\bf 72} (2005) 039902] [hep-ph/0506031].

\bibitem{Blogs5}
  W.-M.~Yao {\it et al.} (Particle Data Group), J. Phys. G {\bf 33}
  (2006) 337. 

\bibitem{Blogs6}
  M.~Ishida and K.~Igi, Phys. Lett. B {\bf 670} (2009) 395
  [arXiv:0809.2424 [hep-ph]]. 

\bibitem{Blogs7}
  M.~Ishida and K.~Igi, Prog. Theor. Phys. Suppl. {\bf 187} (2011) 297.

\bibitem{Blogs8}
  M.~M.~Block and F.~Halzen, Phys. Rev. Lett. {\bf 107} (2011) 212002 
  [arXiv:1109.2041 [hep-ph]].

\bibitem{pdg}  J.~Beringer et al. (Particle Data Group),
  Phys. Rev. D {\bf 86} (2012) 010001.

\bibitem{FLM1}
  M.~Froissart, Phys. Rev. {\bf 123} (1961) 1053.

\bibitem{FLM2}
  A.~Martin, Il Nuovo Cimento {\bf 42A} (1966) 930.

\bibitem{FLM3}
  L.~\L ukaszuk and A.~Martin, Il Nuovo Cimento {\bf 52A} (1967) 122.


\bibitem{Martin} A.~Martin,
  Phys.\ Rev.\ D {\bf 80} (2009) 065013
  [arXiv:0904.3724 [hep-ph]].

\bibitem{WMRS} T.~T.~Wu, A.~Martin, S.~M.~Roy and V.~Singh,
  Phys.\ Rev.\ D {\bf 84} (2011) 025012
  [arXiv:1011.1349 [hep-ph]].

\bibitem{MR} A.~Martin and S.~M.~Roy,
  arXiv:1306.5210 [hep-ph].

\bibitem{GdR} D.~Greynat and E.~de Rafael,
  Phys.\ Rev.\ D {\bf 88} (2013) 034015
  [arXiv:1305.7045 [hep-ph]].

\bibitem{soft-pomeron1}
  L.~L.~Jenkovszky, B.~V.~Struminsky and A.~N.~Wall, Yad. Fiz. {\bf 46} (1987)
  1519.

\bibitem{soft-pomeron2}
  J.~Finkelstein, H.~M.~Fried, K.~Kang and C.-I.~Tang, Phys. Lett. B {\bf 232}
  (1989) 257.

\bibitem{BKYZ}
  G.~Ba\c sar, D.~E.~Kharzeev, H.-U.~Yee, and I.~Zahed, Phys.\ Rev.\ D
  {\bf 85} (2012) 105005 [arXiv:1202.0831 [hep-th]].
  
\bibitem{CGC1}
  E.~Ferreiro, E.~Iancu, K.~Itakura and L.~McLerran, Nucl. Phys. A {\bf 710}
  (2002) 373 
  [hep-ph/0206241].

\bibitem{CGC2}
  L.~Frankfurt, M.~Strikman and M.~Zhalov, Phys. Lett. B {\bf 616}
  (2005) 59 [hep-ph/0412052].

\bibitem{Heisenberg}
  W.~Heisenberg, Zeitschrift f\"ur Physik {\bf 133} (1952) 65.

\bibitem{DGN}
  H.~G.~Dosch, P.~Gauron and B.~Nicolescu, Phys. Rev. D {\bf 67} (2003)
  077501 [hep-ph/0206214].

\bibitem{FMS1}
  D.~A.~Fagundes, M.~J.~Menon and P.~V.~R.~G.~Silva,
  Braz.\ J.\ Phys.\  {\bf 42} (2012) 452
  [arXiv:1112.4704 [hep-ph]].

\bibitem{FMS2}
  D.~A.~Fagundes and M.~J.~Menon,
  Nucl.\ Phys.\ A {\bf 880} (2012) 1
  [arXiv:1112.5115 [hep-ph]].

\bibitem{FMS3}
  D.~A.~Fagundes, M.~J.~Menon and P.~V.~R.~G.~Silva,
  J.\ Phys.\ G {\bf 40} (2013) 065005
  [arXiv:1208.3456 [hep-ph]].

\bibitem{Azimov} Y.~I.~Azimov, Phys. Rev. D
  {\bf 84} (2011) 056012 [arXiv:1104.5314 [hep-ph]].

\bibitem{Nachtmann91}
  O.~Nachtmann, Ann. Phys. {\bf 209} (1991) 436.

\bibitem{pomeron-book}
  S.~Donnachie, G.~Dosch, P.~Landshoff and O.~Nachtmann, {\it Pomeron Physics
    and QCD} (Cambridge University Press, Cambridge, 2002).

\bibitem{DFK}
  H.~G.~Dosch, E.~Ferreira and A.~Kr{\"a}mer, Phys. Rev. D {\bf 50}
  (1994) 1992 [hep-ph/9405237].

\bibitem{Nachtmann97}
  O.~Nachtmann, in {\it Perturbative and Nonperturbative aspects of Quantum
    Field Theory}, edited by H.~Latal and W.~Schweiger (Springer--Verlag,
  Berlin, Heidelberg, 1997) [hep-ph/9609365].

\bibitem{BN}
  E.R.~Berger and O.~Nachtmann, Eur. Phys. J. C {\bf 7} (1999) 459
  [hep-ph/9808320]. 

\bibitem{Dosch}
  H.~G.~Dosch, in {\it At the frontier of Particle Physics -- Handbook of QCD
    (Boris Ioffe Festschrift)}, edited by M.~Shifman (World Scientific, Singapore,
  2001), vol. 2, 1195--1236.
  
\bibitem{LLCM1}
  A.~I.~Shoshi, F.~D.~Steffen and H.~J.~Pirner, Nucl. Phys. A {\bf 709}
  (2002) 131 [hep-ph/0202012]. 
  
\bibitem{analytic1}
  E.~Meggiolaro, Z. Phys. C {\bf 76} (1997) 523 [hep-th/9602104].

\bibitem{analytic2}
  E.~Meggiolaro, Eur. Phys. J. C {\bf 4} (1998) 101 [hep-th/9702186].

\bibitem{analytic3}
  E.~Meggiolaro, Nucl. Phys. B {\bf 625} (2002) 312 [hep-ph/0110069].
  
\bibitem{Meggiolaro05}
  E.~Meggiolaro, Nucl. Phys. B {\bf 707} (2005) 199 [hep-ph/0407084].

\bibitem{GM2009}
  M.~Giordano and E.~Meggiolaro, Phys. Lett. B {\bf 675} (2009) 123
  [arXiv:0902.4145 [hep-ph]]. 

\bibitem{LLCM2}
  A.~I.~Shoshi, F.~D.~Steffen, H.~G.~Dosch and H.~J.~Pirner, Phys. Rev. D {\bf 68}
  (2003) 074004 [hep-ph/0211287].

\bibitem{ILM}
  E.~Shuryak and I.~Zahed, Phys. Rev. D {\bf 62} (2000) 085014
  [hep-ph/0005152]. 

\bibitem{GM2010}
  M.~Giordano and E.~Meggiolaro, Phys. Rev. D {\bf 81} (2010) 074022
  [arXiv:0910.4505 [hep-ph]]. 

\bibitem{JP}
  R.~A.~Janik and R.~Peschanski, Nucl. Phys. B {\bf 565} (2000) 193
  [hep-th/9907177]. 

\bibitem{JP2}
  R.~A.~Janik and R.~Peschanski, Nucl. Phys. B {\bf 586} (2000) 163
  [hep-th/0003059].

\bibitem{Janik}
  R.~A.~Janik, Phys. Lett. B {\bf 500} (2001) 118 [hep-th/0010069].

\bibitem{GP2010}
  M.~Giordano and R.~Peschanski, JHEP 
  {\bf 1005} (2010) 037 [arXiv:1003.2309 [hep-ph]].

\bibitem{GM2008}
  M.~Giordano and E.~Meggiolaro, Phys. Rev. D {\bf 78} (2008) 074510
  [arXiv:0808.1022 [hep-lat]]. 

\bibitem{GM2011a}
  E.~Meggiolaro and M.~Giordano, Prog. Theor. Phys. Suppl. {\bf 187}
  (2011) 200 [arXiv:1010.0914 [hep-lat]].

\bibitem{GM2011b}
  M.~Giordano and E.~Meggiolaro, PoS Lattice {\bf 2011} 155 [arXiv:1110.5188
  [hep-lat]].

\bibitem{BB}
  A.~Babansky and I.~Balitsky, Phys. Rev. D {\bf 67} (2003) 054026
  [hep-ph/0212075]. 

\bibitem{GMM} M.~Giordano, E.~Meggiolaro and N.~Moretti, JHEP 
  {\bf 1209} (2012) 031 [arXiv:1203.0961 [hep-ph]].


\bibitem{reggeon} M.~Giordano,
  JHEP {\bf 1207} (2012) 109
   [Erratum-ibid.\ {\bf 1301} (2013) 021]  
   [arXiv:1204.3772 [hep-ph]].


\bibitem{DR} M.~Rueter and H.~G.~Dosch,
  Phys.\ Lett.\ B {\bf 380} (1996) 177
  [hep-ph/9603214].

\bibitem{crossing1}
  M.~Giordano and E.~Meggiolaro, Phys. Rev. D {\bf 74} (2006) 016003
  [hep-ph/0602143].
\ 
\bibitem{crossing2}
  E.~Meggiolaro, Phys. Lett. B {\bf 651} (2007) 177 [hep-ph/0612307]. 

\bibitem{CT}
  H.~Cheng, E.~Tsai,  Phys.\ Rev.\ D {\bf 36} (1987) 3196.

\bibitem{Weinberg} S.~Weinberg, {\it The Quantum Theory of
  Fields. Vol. 1: Foundations} (Cambridge University Press, Cambridge, 1995).

\bibitem{LSZ1}
  H.~Lehmann, K.~Symanzik and W.~Zimmermann,
  Nuovo Cim.\  {\bf 1} (1955) 205.

\bibitem{LSZ2}
  H.~Lehmann, K.~Symanzik and W.~Zimmermann,
  Nuovo Cim.\  {\bf 6} (1957) 319.

\bibitem{unitarity1}
  U.~Amaldi, M.~Jacob, G.~Matthiae, Ann. Rev. Nucl. Part. Sci. {\bf 26} (1976)
  385.

\bibitem{unitarity2}
  R.~Castaldi and G.~Sanguinetti, Ann. Rev. Nucl. Part. Sci. {\bf 35} (1985)
  351.

\bibitem{unitarity3}
  M.~M.~Block and R.~N.~Cahn, Rev. Mod. Phys. {\bf 57} (1985) 563.


\bibitem{Eden} R.~J.~Eden,
  Rev.\ Mod.\ Phys.\  {\bf 43} (1971) 15.

\bibitem{AKM}
  G.~Auberson, T.~Kinoshita and A.~Martin,
  Phys.\ Rev.\ D {\bf 3} (1971) 3185.

\bibitem{LambertW}  R.~M.~Corless, G.~H.~Gonnet, D.~E.~Hare,
  D.~J.~Jeffrey, and D.~E.~Knuth,   
  Adv.\ Comput.\ Math.\ {\bf 5} (1996) 329.

\bibitem{TT1}
  S.~M.~Troshin, N.~E.~Tyurin, Int. J. Mod. Phys. A
  {\bf 22} (2007) 4437 [hep-ph/0701241].

\bibitem{TT2}
  S.~M.~Troshin, N.~E.~Tyurin, Phys. Lett. B {\bf 316} (1993) 175
  [hep-ph/9307250]. 

\bibitem{BDdD} I.~Bautista and J.~Dias de Deus,
  Phys.\ Lett.\ B {\bf 718} (2013) 1571
  [arXiv:1212.1764 [nucl-th]].

\bibitem{CN}  T.~Cs\"org\H o and F.~Nemes,
Int. J. Mod. Phys. A {\bf 29} (2014) 1450019
  [arXiv:1306.4217 [hep-ph]].


\bibitem{GNL} P.~Gauron, B.~Nicolescu and E.~Leader,
  Nucl.\ Phys.\ B {\bf 299} (1988) 640.

\bibitem{ABBW}
  G.~Audi, O.~Bersillon, J.~Blachot and A.~H.~Wapstra,
  Nucl.\ Phys.\ A {\bf 729} (2003) 3.


\bibitem{MP} C.~J.~Morningstar and M.~J.~Peardon,
  Phys.\ Rev.\ D {\bf 60} (1999) 034509
  [hep-lat/9901004].

\bibitem{Ky}
  Y.~Chen, A.~Alexandru, S.~J.~Dong, T.~Draper, I.~Horv\'ath, F.~X.~Lee,
  K.~F.~Liu, N.~Mathur, C.~Morningstar, M.~Peardon,
  S.~Tamhankar, B.~L.~Young, and J.~B.~Zhang, 
  Phys.\ Rev.\ D {\bf 73} (2006) 014516
  [hep-lat/0510074].

\bibitem{Meyer}
  H.~B.~Meyer, 
  hep-lat/0508002.

\bibitem{GILMRRR}  E.~Gregory, A.~Irving, B.~Lucini, C.~McNeile,
  A.~Rago, C.~Richards and E.~Rinaldi, 
  JHEP {\bf 1210} (2012) 170
  [arXiv:1208.1858 [hep-lat]].


\bibitem{Moriond} E.~Meggiolaro, M.~Giordano and N.~Moretti,
  arXiv:1304.3297 [hep-ph].

\bibitem{EMN}
  C.~Ewerz, M.~Maniatis and O.~Nachtmann,
Annals Phys.\  {\bf 342} (2014) 31
  [arXiv:1309.3478 [hep-ph]].



\end{thebibliography}
\end{document}